\documentclass[a4paper,12pt]{article}
\usepackage{amsfonts,amsthm,amsmath,amssymb}
\usepackage{graphicx}
\numberwithin{equation}{section}
\usepackage{ascmac}

\begin{document}

\newtheorem{definition}{Definition}[section]
\newcommand{\be}{\begin{equation}}
\newcommand{\ee}{\end{equation}}
\newcommand{\bea}{\begin{eqnarray}}
\newcommand{\eea}{\end{eqnarray}}
\newcommand{\LE}{\left[}
\newcommand{\R}{\right]}
\newcommand{\nn}{\nonumber}
\newcommand{\Tr}{\text{Tr}}
\newcommand{\N}{\mathcal{N}}
\newcommand{\G}{\Gamma}
\newcommand{\vf}{\varphi}
\newcommand{\LL}{\mathcal{L}}
\newcommand{\Op}{\mathcal{O}}
\newcommand{\HH}{\mathcal{H}}
\newcommand{\arctanh}{\text{arctanh}}
\newcommand{\up}{\uparrow}
\newcommand{\down}{\downarrow}
\newcommand{\ket}[1]{\left| #1 \right>}
\newcommand{\bra}[1]{\left< #1 \right|}
\newcommand{\ketbra}[1]{\left|#1\right>\left<#1\right|}
\newcommand{\rd}{\partial}
\newcommand{\de}{\partial}
\newcommand{\ba}{\begin{eqnarray}}
\newcommand{\ea}{\end{eqnarray}}
\newcommand{\db}{\bar{\partial}}
\newcommand{\we}{\wedge}
\newcommand{\ca}{\mathcal}
\newcommand{\lr}{\leftrightarrow}
\newcommand{\f}{\frac}
\newcommand{\s}{\sqrt}
\newcommand{\vp}{\varphi}
\newcommand{\hvp}{\hat{\varphi}}
\newcommand{\tvp}{\tilde{\varphi}}
\newcommand{\tp}{\tilde{\phi}}
\newcommand{\ti}{\tilde}
\newcommand{\ap}{\alpha}
\newcommand{\pr}{\propto}
\newcommand{\mb}{\mathbf}
\newcommand{\ddd}{\cdot\cdot\cdot}
\newcommand{\no}{\nonumber \\}
\newcommand{\la}{\langle}
\newcommand{\lb}{\rangle}
\newcommand{\ep}{\epsilon}
 \def\we{\wedge}
 \def\lr{\leftrightarrow}
 \def\f {\frac}
 \def\ti{\tilde}
 \def\ap{\alpha}
 \def\pr{\propto}
 \def\mb{\mathbf}
 \def\ddd{\cdot\cdot\cdot}
 \def\no{\nonumber \\}
 \def\la{\langle}
 \def\lb{\rangle}
 \def\ep{\epsilon}
\newcommand{\mcl}{\mathcal}
 \def\g{\gamma}
\def\tr{\text{tr}}

\begin{titlepage}
\thispagestyle{empty}

\begin{flushright}
EFI-17-7\\
\end{flushright}
\bigskip
\begin{center}
\noindent{\large \textbf{Correspondence between Entanglement Growth and \\ Probability Distribution of Quasi-Particles}}\\
\vspace{2cm}

Masahiro Nozaki $^{a}$ and Naoki Watamura $^{b}$ \\

\vspace{1cm}

{\it $^{a}$Kadanoff Center for Theoretical Physics, University of Chicago,\\
Chicago, Illinois 60637, USA \\}

{\it
$^{b}$Department of Physics
Nagoya University, Nagoya 464-8602, Japan\\}

\vskip 4em
\end{center}

\begin{abstract}
	We study the excess of (Renyi) entanglement entropy in various free field theories for the locally excited states defined by acting with local operators on the ground state. It is defined by subtracting the entropy for the ground state from the one for the excited state. 
	Here the spacetime dimension is greater than or equal to 4.
	We find a correspondence between entanglement and a probability.
	The probability with which a quasi-particle exists in a subregion gives the excess of the entropy.
	We also propose a toy model which reproduces the excess in the replica method.
	In this model, a quasi-particle created by a local operator propagates freely and its probability distribution gives the excess.
\end{abstract}
\end{titlepage}

\tableofcontents
\section{Introduction and Summary}
(R\'enyi) entanglement entropy is expected to be a useful tool to diagnose the non-equilibrium physics such as thermalization, creation and evaporation of the black hole \cite{C1,C2,C3,MM1,MM2,HL1,HL2,HM,LP}.
Currently many researchers try to construct quantum gravity by using  entanglement in the theory living on the boundary \cite{r,t1,sw1,sw2,tt2,r2,dh,t3,oog,MI}. 

It is important that the fundamental properties of quantum entanglement is studied in this trial. 
In this paper, we study its dynamical property. Before explaining our results in summary, we explain the results which has been obtained in various protocol. 
The dynamics of entanglement has been studied by measuring (R$\acute{e}$nyi) entanglement entropy in various protocols. One of the protocol is called global quenches where a parameter of  Hamiltonian is suddenly changed \cite{C1, C2}. 
The time evolution of entanglement entropy in 2 dimensional conformal field theories (CFTs) is well-known. 
We assume that  Hamiltonian is changed at $t=0$ and entanglement entropy is measured at $t' (>0)$.
Here $l (>0)$ is the subsystem size. If $t' < l/2$, entanglement entropy linearly increases with $t$.
If $t' \ge l/2$ entanglement entropy stops to increase and is proportional to the subsystem size $l$ (volume law).
Its time evolution in $ t' < l/2$ is interpreted in terms of the relativistic propagation of quasi-particles which are entangled.
Its volume law in $t' \ge l/2$ comes from entanglement between quasi-particles. 
Recently the time evolution of entanglement entropy for global quenches is studied in higher dimensional CFTs and holographic field theories \cite{MM1,MM2,HL1,HL2,HM,LP}.

Another protocol is called local quenches.
 Hamiltonian is deformed locally at $t$. A well-known result in these quenches is the time evolution of entanglement entropy in the $2$-dimensional CFTs\cite{CL}.
The time evolution of entanglement entropy can be interpreted in terms of the relativistic propagation of quasi-particles even in local quenches. 
A holographic model of these quenches is proposed in \cite{TL,UL,RL} and the author in \cite{WX} discusses the relation between global and local quenches. 

Recently, entanglement entropy has been studied in more general quenches where the state is not suddenly excited but continuously excited with respect to $t$ \cite{Das, Zu1, Zu2}.

In a simpler protocol,  locally excited states are defined not by deforming  Hamiltonian but by acting with local operators on the ground state.
In the articles \cite{MN1,MN2,MN4} the time evolution of (R\'enyi) entanglement entropy for those states in various free field theories has been studied. 
The excess of (R\'enyi) entanglement entropy $\Delta S^{(n)}_A$ is defined by subtracting the entropy for the ground state from the one for the excited state because the ground-state entropy is static quantity.
The author in \cite{NonR} studied $\Delta S^{(n)}_A$ in a non-relativistic system. 
The time evolution of $\Delta S^{(n)}_A$ can be qualitatively interpreted in terms of the relativistic propagation of the quasi-particles created by a local operator. Furthermore, the late-time value of the entropy can be quantitatively interpreted in terms of quasi-particles. Its reduced density matrix can be given by their probability distribution.
Even in the interacting and holographic CFTs \cite{Alc, TKO1,KP1,SH,TKO2,SJ,MN4,TH,Cap1,Bj}, the time evolution of these entropies can be qualitatively interpreted in terms of their relativistic propagation. 
However in the late time limit which will be precisely explained later, their behavior depends on the theory. In the solvable theories such as minimal models, they are given by the quantum dimension of an inserted local operator. On the other hand, in the holographic theories, the entropy increases logarithmically with $t$. 

{\bf Summary}

In this paper we study $\Delta S^{(n)}_A$ in the various free field theories (in particular, free massless scalar theories and free Maxwell theories). We find that its time evolution for any $t$ is given by  (R\'enyi) entanglement entropy whose reduced density matrix is given by the probability distribution of the quasi-particle created by the local operator.
Here we assume that the spacetime dimension is greater than or equal to $4$.  
If the subsystem $A$ is given by the half of the total system, a kind of quasi-particle is included in $A$ with the probability $P_A$ which can be given by a propagator.
Not only the late time values but also the whole  time evolution of $\Delta S^{(n)}_A$ can be equal to  (R\'enyi) entanglement entropy whose reduced density matrix is given by the probability distribution of the quasi-particle created by the local operator.

We propose a toy model where quasi-particles created by local operators freely propagate at the speed of light.
$\Delta S^{(n)}_A$ is given by an ``entropy" with their probability distribution. In section 4, we will explain its definition.
By using this model, we estimate $\Delta S^{(n)}_A$ for a more complicated shaped subregion than the ones discussed previously. The excess of mutual information $\Delta I_{A, B}$ in some cases is estimated. 

{\bf Organization}

This paper is organized as follows.
In section $2$, we explain how to compute $\Delta S^{(n)}_A$ in the replica method.
In section $3$, we explain the correspondence between the existence probability and propagators in the replica method.
In section $4$, we propose a toy model, in which an operator creates a quasi-particle.
We show that $\Delta S^{(n)}_A$ is given by the entropy with its probability distribution. $\Delta S^{(n)}_A$ and $\Delta I_{A, B}$ in some cases are estimated by this entropy.

\section{Entanglement Entropy in the Replica Method}


\subsection{The space decomposition.}
Here we are dealing with quantum field theories (QFTs) with $d+1$ dimensional Lorentzian spacetime\footnote{The theories are put on even dimensional Minkowski spacetime with signature $g_{\mu \nu}= \text{diag}(-1,1,1,\cdots)$ in the following section.}. 


\begin{figure}[tbh]
	\centering
	\includegraphics[width=60mm]{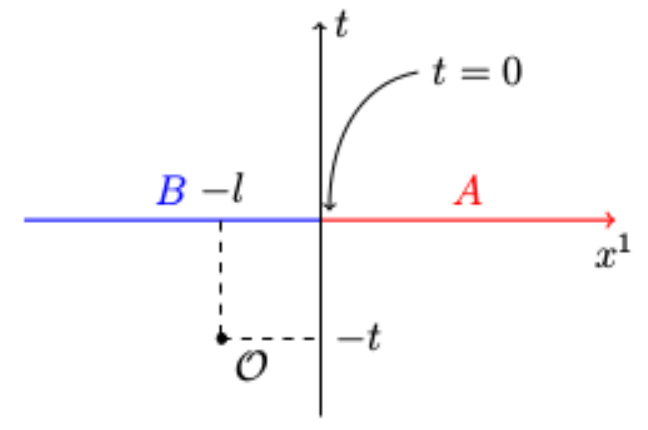}
	\caption{The total Space divided into two subspaces $A$ and $B$ at $t=0$.}
	\label{fig:spdc}
\end{figure}
A definition of (R\'enyi) entanglement entropy $S^{(n)}_A$ in QFTs is as follows. 
The total Hilbert space is geometrically divided into $A$ and $B$. Here it is done at $t=0$ in order to measure $S^{(n)}_A$ at $t=0$.
In this paper $A$ is defined by $x_1 \ge 0$ and $B$ is its complement as in Fig.\ref{fig:spdc}. A reduced density matrix $\rho_A$ for $A$ is defined by tracing out the degrees of freedom in $B$:
\be
\rho_A :=\Tr_B \rho,
\ee 
where $\rho$ is a given density matrix.
Its (R\'enyi) entanglement entropy is defined by
\be
S^{(n)}_A= \f{1}{1-n}\log{\left[\Tr_A\left(\rho_A\right)^n\right]}.
\ee

\subsection{Locally Excited State and $\Delta S_A$}
Our interest is to study the dynamics of quantum entanglement. We define the excess of (R\'enyi) entanglement entropy $\Delta S^{(n)}_A$ by subtracting the entropy for the ground state $ S^{(n),G}_A$ from the one for an excited state $S^{(n), EX}_A$ since the ground-state entropy does not depend on time:
\begin{align}\label{EREE}
	\Delta S_A^{(n)} := S_A^{(n), \mbox{\tiny EX}} - S_A^{(n), \mbox{\tiny G}}.
\end{align}

A given excited state in this paper is a locally exited state:
\be
\ket{\Psi}=\mathcal{N}\mathcal{O}(-t, -l, {\bf x})\ket{0},
\ee
 where the local operator is located at $(t, x_1, {\bf x} )=(-t, -l, {\bf x})$ and $\mathcal{N}$ is a normalization constant (Fig.\ref{fig:spdc}).
The coordinate in Mikowski spacetime is written by $(t, x_1, {\bf x})$, where ${\bf x}=(x_2, \cdots, x_{d})$.
In the following subsection, we will explain how to compute $\Delta S^{(n)}_A$ for the locally excited state in the replica method.

\begin{figure}[tb]
	\centering
	\begin{tabular}{cc}
		\includegraphics[width=60mm]{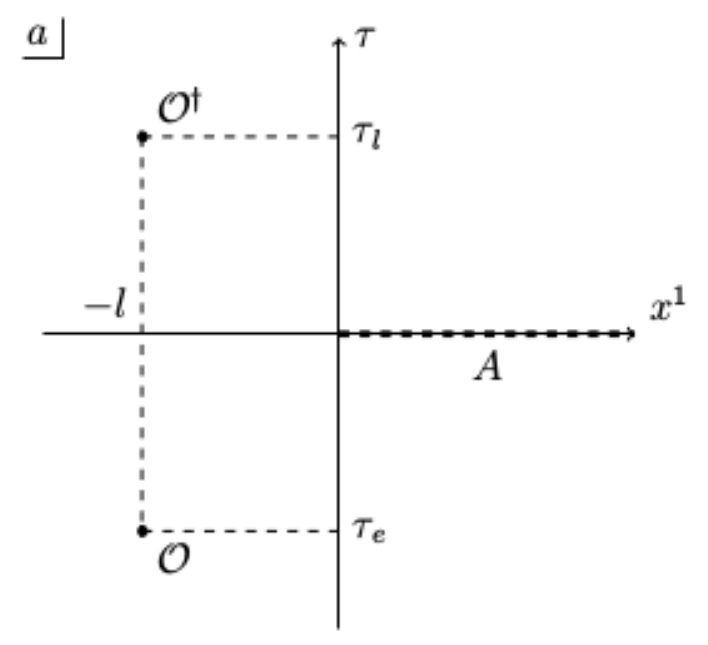}
		&
		\includegraphics[width=60mm]{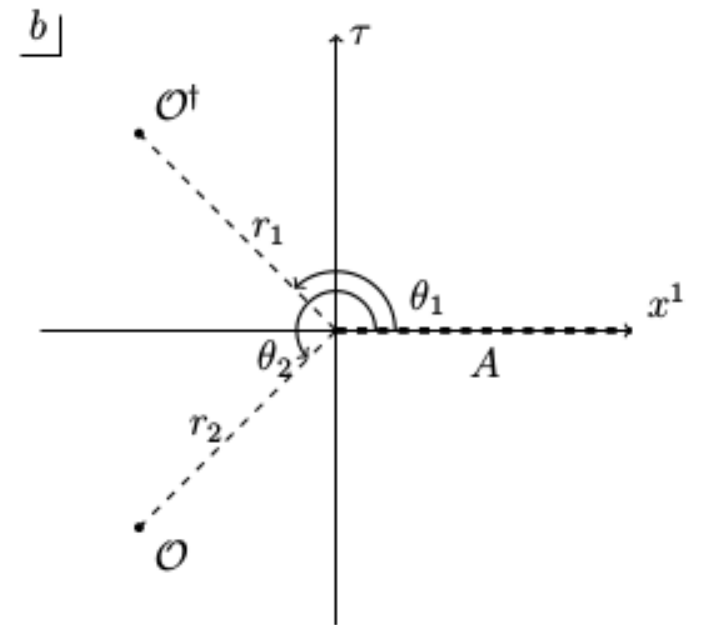}
	\end{tabular}
	\caption{Operator insertion points before taking the analytic continuation in a: $(\tau, x^1)$ and b:$(r, \theta)$.}
	\label{fig:insertionpoints}
\end{figure}


\subsection{The Replica Method}

Let's explain how to compute the excess of (R\'enyi) entanglement entropy in the replica method. 
A given density matrix in $d+1$ dimensional Euclidean space is 
\begin{align} \label{mex}
	\rho^{\mbox{\tiny ex}} &= {\cal N}^2 {\cal O} (\tau_e, -l, {\bf x}) | 0 \rangle \langle 0 | {\cal O}^\dagger (\tau_l, -l, {\bf x}) .
\end{align}
where $\mathcal{N}$ is a normalization constant and the coordinate in the space is $(\tau , x_1, {\bf x})$. ${\bf x}=\{x_2,\cdots, x_d\}$.
The density matrix can be schematically interpreted as in Fig.\ref{fig:insertionpoints}. In the figure, a local operator $\mathcal{O}$ is located at $(\tau_e, -l, {\bf x})$ and its ``conjugate" operator $\mathcal{O}^{\dagger}$ is at $(\tau_l, -l, {\bf x})$.

Even in Euclidean space, the excess of (R\'enyi) entanglement entropy is defined by 
\begin{align} \label{fm}
	\Delta S_{A, Eu}^{(n)} := S_{A, Eu}^{(n), \mbox{\tiny EX} } - S_{A, Eu}^{(n), \mbox{\tiny G}}.
\end{align}
where $S_{A, Eu}^{(n)}$ is (R\'enyi) entanglement entropy defined in Euclidean space. $S_{A, Eu}^{(n), \mbox{\tiny EX} }$ and $S_{A, Eu}^{(n), \mbox{\tiny G}}$ are the entropies for $\rho^{\mbox{\tiny ex}}$ and the ground state, respectively.  The entropy in Euclidean space is just written by $S^{(n)}_A$ in the following.
The replica method is well known, and we recommend \cite{Rep} for further reading.
For convenience we give here a brief description.

Let's compute $S_A^{(n)}$ for the ground state in the replica method.
In Euclidian QFTs, the wave functional at $\tau=0$ of a vacuum state $\Psi^{\mbox{\tiny vac}}[\phi_0(x)]$ can be described in the path-integral form as
\begin{align}
	\Psi^{\mbox{\tiny vac}} [ \phi_0 (x) ] &= \frac{1}{\left( Z_1^{\mbox{\tiny vac}} \right)^{\frac 12}} \int_{\phi(\tau = - \infty)}^{\phi(\tau =0,x) = \phi_0(x)} {\cal D}\phi ~ e^{-S[\phi]}, \label{vac}
\end{align}
where $x$ is the space coordinate $x = \{x_1, {\bf x}\}$ and $Z^{\mbox{\tiny vac}}_1$ is the partition function of the vacuum (on the spacetime ${\mathbb R}^{d+1}$).
With this expression, we can rewrite the density matrix of vacuum $\rho^{\mbox{\tiny vac}}[\phi_-(x),\phi_+(y)]$ as
\begin{align}
	\begin{split}
		\rho^{\mbox{\tiny vac}} [\phi_-(x), \phi_+(y)] 
		&= \Psi^{\mbox{\tiny vac}}[\phi_-(x)] \Psi^{\mbox{\tiny vac} \dagger}[\phi_+(y)] \\
		&= \frac{1}{Z_1^{\mbox{\tiny vac}}} \int_{\phi_1(t=-\infty)}^{\phi_1(\tau=0_-,x) = \phi_-(x)} {\cal D} \phi_1 ~ e^{-S[\phi_1]} ~ \int_{\phi_2(\tau=0_+,y) =\phi_+(y)}^{\phi_2(t=\infty)}{\cal D}\phi_2 ~ e^{-S[\phi_2]},
	\end{split} \label{eq:defdensmtx}
\end{align}
where $\phi_\pm$ are the boundary conditions at $\tau=0_{\pm}$($0_{\pm}=0 \pm \delta (\ll1)$.).

\begin{figure}
	\centering
	\begin{tabular}{cc}
		\includegraphics[width=60mm]{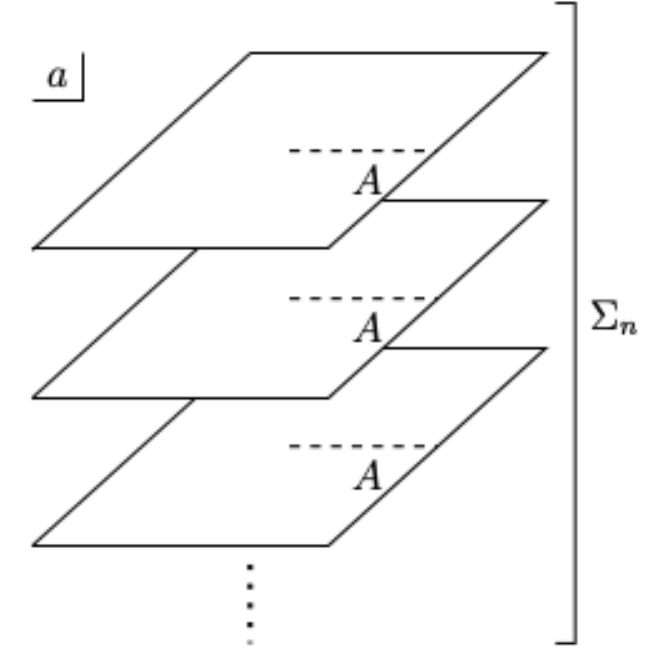}
		&
		\includegraphics[width=60mm]{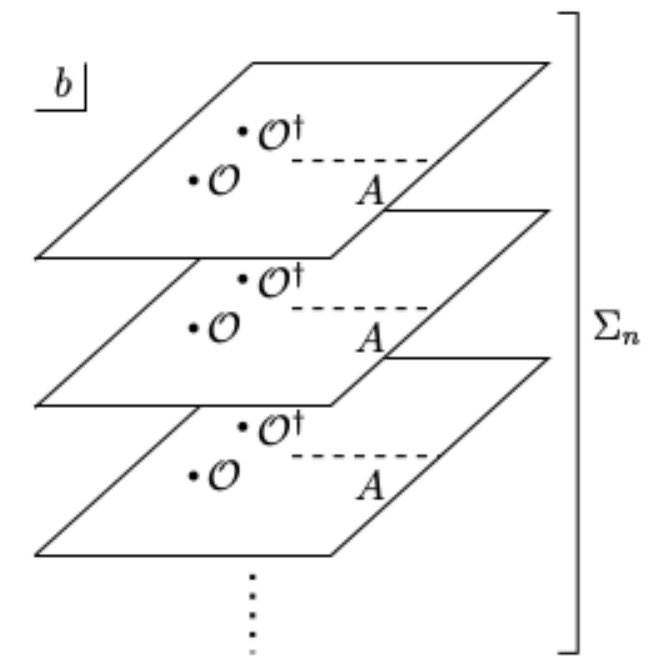}
	\end{tabular}
	\caption{Sketch of a: $n$-sheeted Riemann surface, b: $n$-sheeted Riemann surface with operator insertions.}
	\label{fig:nsigma}
\end{figure}

The reduced density matrix $\rho^{\mbox{\tiny vac}}_A$ is defined by tracing out the degrees of freedom in the region $B$.
Since $A$ is the half of total space,  the degrees in $A$ and $B$ can be defined by
\be
\begin{split}
\phi(\tau, x):=\begin{cases}
\phi^{A}(\tau, x) & x_1 >0, \\
\phi^{B}(\tau,x ) & x_1 <0. \\
\end{cases}
\end{split}
\ee
Then the matrix $\rho^{\mbox{\tiny vac}}_A[\phi^A_-(x), \phi^A_+(y)]$ is given by 
\begin{align}
	\rho^{\mbox{\tiny vac}}_A \left[ \phi^A_-(x), \phi^A_+(y) \right] &= \int \mathcal{D}\phi^B_+ \int \mathcal{D} \phi^B_- \delta(\phi^B_+(x)-\phi^B_-(x))\rho^{\mbox{\tiny vac}} [\phi_-(x), \phi_+(y)] 
\end{align}
Finally, $\Tr_A\left(\rho^{\mbox{\tiny vac}}_A\right)$ is 
\begin{align}\label{nrv}
	\begin{split}
		\Tr_A \left( \rho_A^{\mbox{\tiny vac}} \right)^n
		&= \frac{1}{\left( Z_1^{\mbox{\tiny vac}} \right)^n} \int_A \left[ \prod_{i=1}^{i=2n} {\cal D} \phi_i \right] \rho_A^{\mbox{\tiny vac}}[\phi_1,\phi_2] \delta(\phi_2 - \phi_3) \rho_A^{\mbox{\tiny vac}}[\phi_3,\phi_4] \cdots \rho_A^{\mbox{\tiny vac}}[\phi_{2n-1},\phi_{2n}] \delta(\phi_{2n-1} - \phi_{2n}) \\
		&= \frac{1}{\left( Z_1^{\mbox{\tiny vac}} \right)^n} \int_{\Sigma_n} {\cal D} \phi ~ e^{-S_n[\phi]}
	\end{split}
\end{align}
where the integrals in the first line are performed over the region $A$ of each Riemann sheet. $\Sigma_n$ is a $n$-sheeted Riemann sheet described in Fig.\ref{fig:nsigma} and $S_n$ is an action defined on $\Sigma_n$.

Let's compute (R\'enyi) entanglement entropy for the state in (\ref{mex}). The matrix in (\ref{mex}) can be
written by 
\begin{align}
	\rho^{\mbox{\tiny ex}} &= \frac{1}{Z^{\mbox{\tiny ex}}_1} {\cal O}( \tau_e , -l) | 0 \rangle \langle 0 | {\cal O}^\dagger ( \tau_l , -l)
\end{align}
We rewrite the coordinates $(\tau, x^1)$ into polar coordinates $(r, \theta)$ as in Fig. \ref{fig:insertionpoints} .
$\Tr_A \rho_A^{\mbox{\tiny ex}}$ gives
\begin{align}\label{nre}
	\Tr_A \left( \rho_A^{\mbox{\tiny ex}} \right)^n &= \frac{1}{\left( Z_1^{\mbox{\tiny ex}} \right)^n} \int_{\Sigma_n} {\cal D}\phi ~ \prod_{i=1}^{n} \left( {\cal O}^\dagger (r_1,\theta_1^{(i)}) ) {\cal O} ( r_2, \theta_2^{(i)} ) \right) e^{-S_n[\phi]}
\end{align}
where $( r_{1,2}, \theta_{1,2}^{(i)} )$ are the insertion points of local operators ${\cal O}^\dagger$ and ${\cal O}$ on the $i$-th Riemann sheet as it is described in Fig.\ref{fig:nsigma} b.
$Z^{\mbox{\tiny ex}}_1 = \int_{\Sigma_1} {\cal D}\phi ~ \left( {\cal O}^\dagger (r_1,\theta_1) ) {\cal O} ( r_2, \theta_2 ) \right) e^{-S}$ is introduced for the normalization.

After substituting (\ref{nrv}) and (\ref{nre}) into (\ref{fm}), $\Delta S^{(n)}_A$ in the replica method is 
\begin{align}
	\begin{split}
		\Delta S_A^{(n)} &:= S_A^{(n), \mbox{\tiny ex}} - S_A^{(n), \mbox{\tiny vac}} \\
		&= \frac{1}{1-n} \left( \log \mbox{tr}_A \left( \rho_A^{\mbox{\tiny ex}} \right)^{n} - \log \Tr_A \left( \rho_A^{\mbox{\tiny vac}} \right)^n \right)\\
		&= \frac{1}{1-n} \log {\left[\frac{\langle \prod_{i=1}^{n} \left( {\cal O}^\dagger (r_1,\theta_1^{(i)}) ) {\cal O} ( r_2, \theta_2^{(i)} ) \right) \rangle_{\Sigma_n}}{\left(\left \langle {\cal O}^\dagger (r_1,\theta_1)  {\cal O} ( r_2, \theta_2 )  \right \rangle_{\Sigma_1} \right)^n}\right]},
	\end{split}
\end{align}
where $\langle {\cal O}(x) {\cal O}^\dagger(y) \cdots \rangle_{\Sigma_N}$ is the correlation function on the $n$-sheeted Riemann surface.

\subsection{Analytic continuation to the real time}

In this method, the $2n$-point function of $\mathcal{O}$ in $\Sigma_n$ and the $2$-point function of $\mathcal{O}$ in $\Sigma_1$ give $\Delta S^{(n)}_A$ in Euclidean spacetime.
In order to study the dynamics of entanglement in Minkowski spacetime, we perform the analytic continuation to the real time as in the articles \cite{MN1,MN2,MN4,TKO1,KP1,SH,MN5}.
The analytic contiunation to the real time is done by 
\begin{align}
	\begin{split}
		\tau_l &= \epsilon - i t\\
		\tau_e &= - \epsilon - i t
	\end{split}
\end{align}
where $\epsilon$ acts as a smearing parameter which keeps the norm of the locally excited state finite. During the calculation, we keep $\epsilon$ finite, but in the end we take the limit $\epsilon \rightarrow 0$. 
Note that in Maxwell theory, the fields also change as
\begin{align}
	\begin{split}
		A_\tau = i A_t \\
		\partial_\tau = i \partial_t
	\end{split}
\end{align}
due to covariance.

\section{Probability and Propagator}
In this section, we study $\Delta S^{(n)}_A$ for locally excited states in the replica method. The spacetime dimensions are assumed to be more than or equal to $4$.  
We explain the correspondence between an analytic-continued propagator and a probability.
\subsection{$\Delta S^{(n)}_A$ in free field theories}
In \cite{MN1,MN2,MN4,TKO1,KP1,SH,MN5}, the time evolution of $\Delta S^{(n)}_A$ in the limit $\epsilon \rightarrow 0$ is studied. In free field theories, the leading term of $\Delta S^{(n)}_A$ does not depend on $\epsilon$ and it is finite. If the late time limit $(t \rightarrow \infty)$ is taken, $\Delta S^{(n)}_A$ is given by (R$\acute{e}$nyi) entanglement entropy whose reduced density matrix is given by an effective reduced density matrix $\rho^{e}_A$:
\be
  \Delta S^{(n)}_A =\f{1}{1-n}\log{[\Tr \left(\rho^{e}_A\right)^n ]},
\ee
where $\Tr\rho^{e}_A =1$.
The density matrix is evaluated by quasi-particles which obey the late time algebra as explained in the following subsection. Therefore, the late time values of $\Delta S^{(n)}_A$ are given by entanglement of quasi-particles. 
\subsubsection{$\rho_A^{e}$ in the Late Time Limit}
Let's explain a quasi-particle picture in the late time limit and the late time algebra which the particle obeys in a simple case. For simplicity, we consider $4$ dimensional massless free scalar field theory. The given local state is 
\be \label{exam}
\ket{\Psi_s}= \mathcal{N} \phi(-t,-l, {\bf x})\ket{0},
\ee 
where $\mathcal{N}$ is determined so that $\left\langle \Psi_s\big{|}\Psi_s \right\rangle=1$ and the operator $\phi$ is included in $B$.

Before explaining the particle picture in the late time limit and the late time algebra, we explain the time evolution of $\Delta S^{(n)}_A$.
The time evolution of $\Delta S_{A}^{(n)}$ for (\ref{exam}) can be qualitatively interpreted in terms of the relativistic propagation of quasi-particle. 

The time evolution of $\Delta S^{(n)}_A$ has three processes.
In $t\le l$, $\Delta S^{(n)}_A$ vanishes.
It increases in $t>l$.
After taking the limit $t \rightarrow \infty$, $\Delta S^{(n)}_A$ approaches a constant.

\begin{figure}[htbp]
  \begin{center}
    \begin{tabular}{c}

      \begin{minipage}{0.33\hsize}
        \begin{center}
          \includegraphics[clip, width=4.5cm]{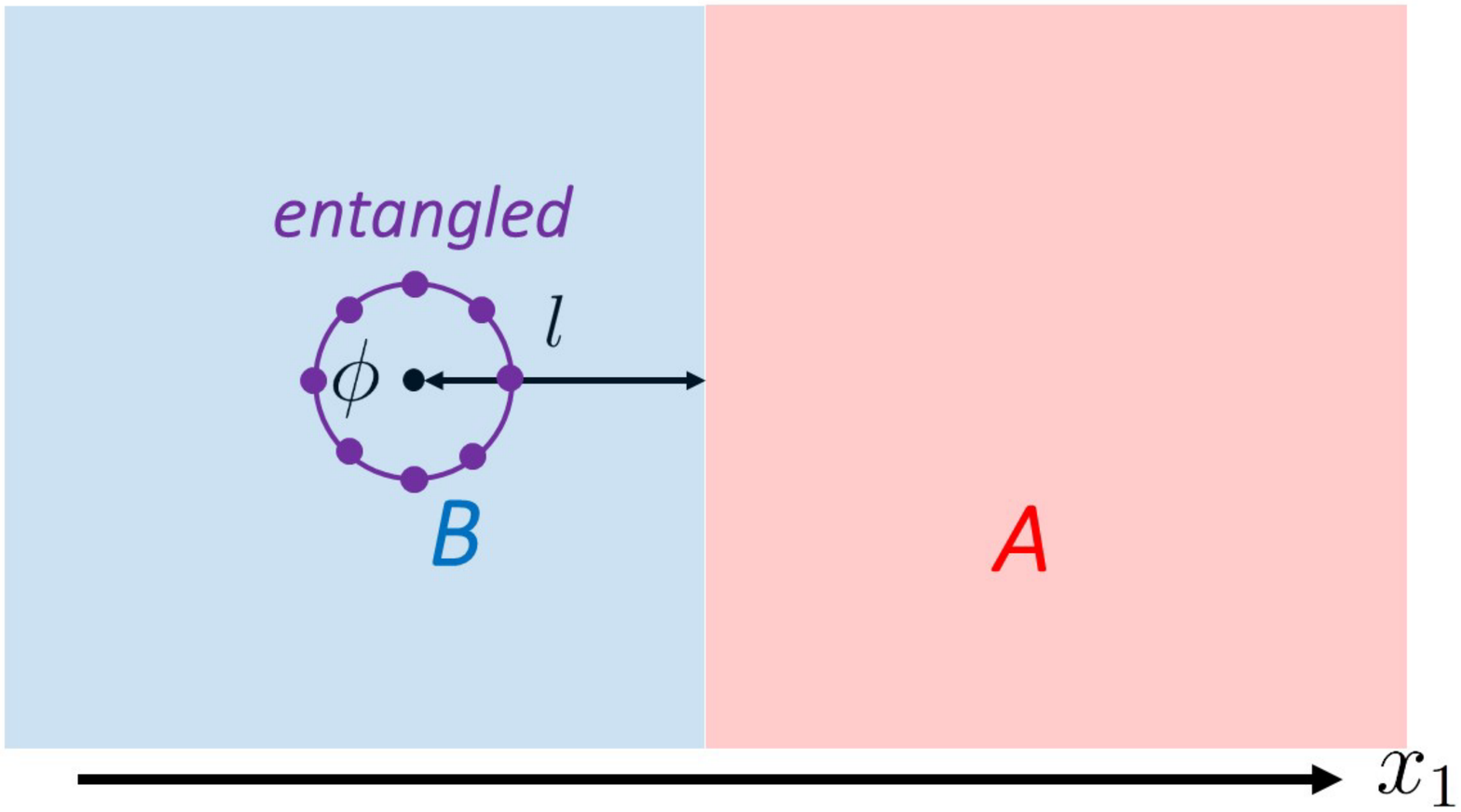}
          \hspace{1.6cm} (a) $l\ge t>0$.
        \end{center}
      \end{minipage}

      \begin{minipage}{0.33\hsize}
        \begin{center}
          \includegraphics[clip, width=4.5cm]{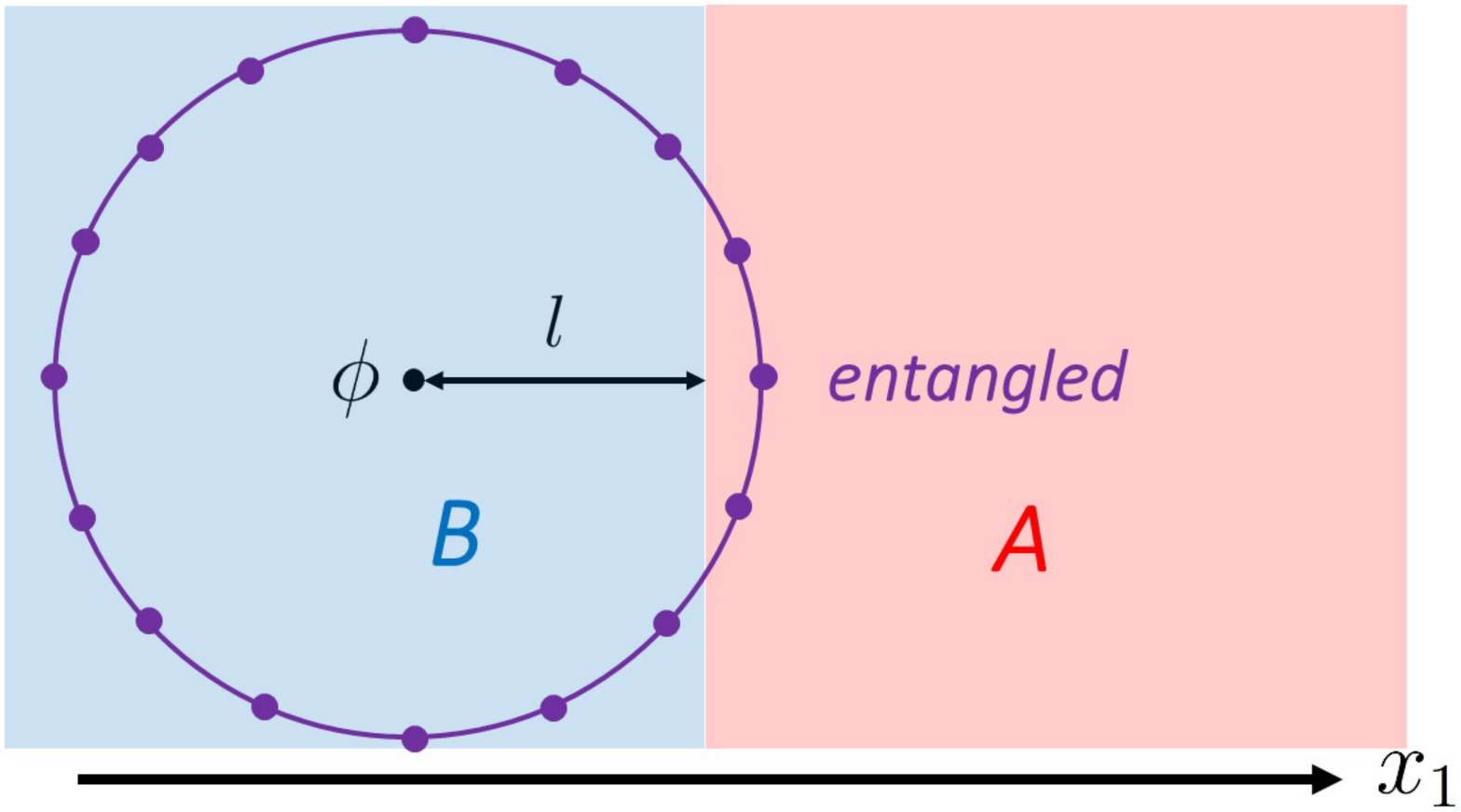}
          \hspace{1.6cm} (b) $t>l>0$.
        \end{center}
      \end{minipage}

      \begin{minipage}{0.33\hsize}
        \begin{center}
          \includegraphics[clip, width=4.5cm]{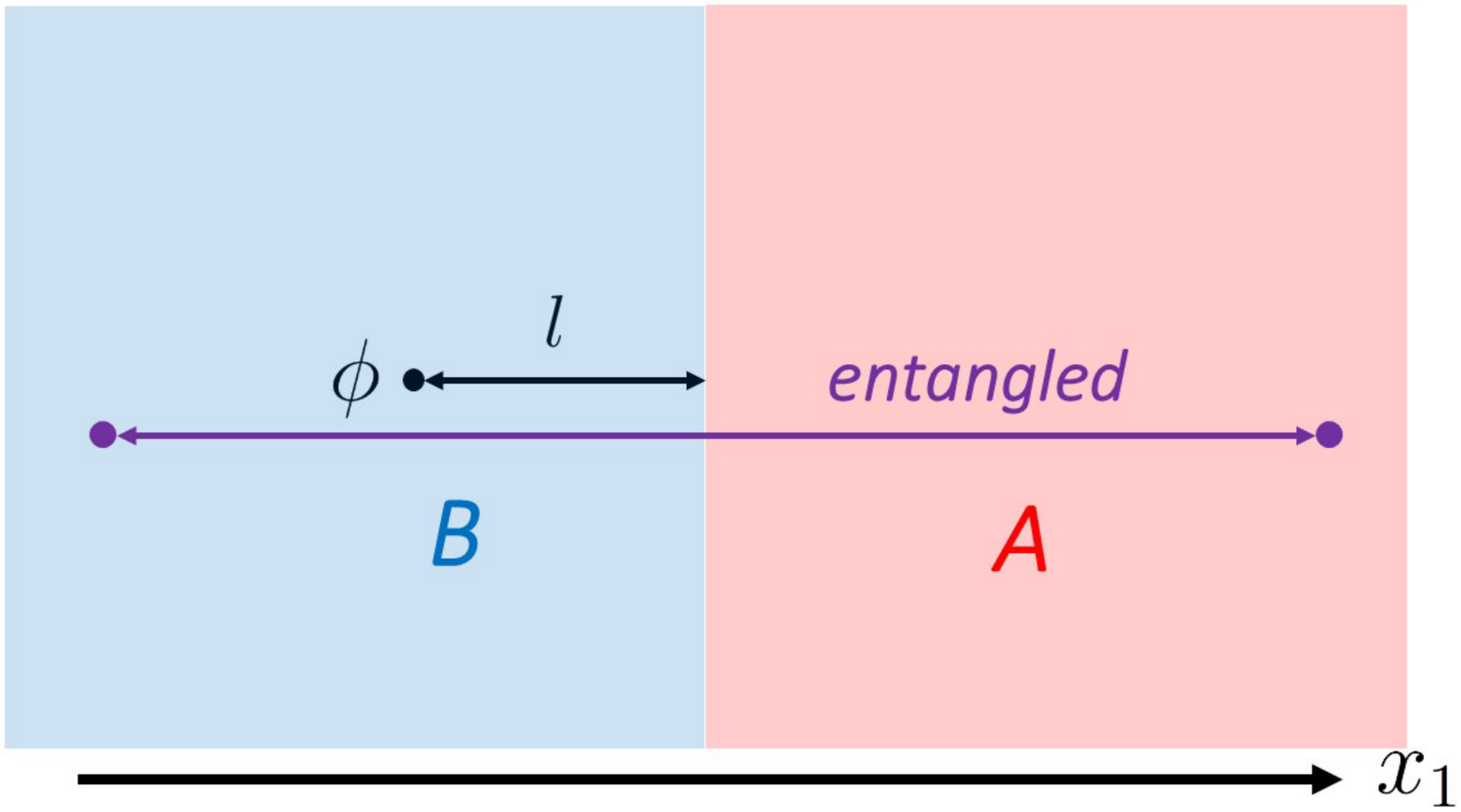}
          \hspace{1.6cm} (c) The late time limit.
        \end{center}
      \end{minipage}

    \end{tabular}
    \caption{A quasi-particle picture. At $t=-t$, quasi-particles appear at the point where $\phi$ is located. (a) shows that all of them are included in $B$ in $l \ge t>0$. Therefore entanglement between them can not contribute to $\Delta S^{(n)}_A$.  (b) shows that their entanglement can contribute to $\Delta S^{(n)}_A$ in $t>l>0$ because some of them are included in $A$. In the late time, it can be interpreted in terms of entanglement between two quasi-particles.}
    \label{fig:lena}
  \end{center}
\end{figure}

In a quasi-particle picture, an entangled group is created at $ (t, x_1, {\bf x})=(-t, -l, {\bf x})$ and spherically propagates at the speed of light. The group is constructed of the quasi-particles entangled each other. In $t \le l$ the group is included in $B$(Fig.\ref{fig:lena}(a)). Then the entanglement between the particles can not contribute to $\Delta S^{(n)}_A$.
If $t>l$, some of them are included in $B$ and entanglement between them can contribute to $\Delta S^{(n)}_A$.
In the late time limit, the particles included in $A$ can not come out of $A$.
Their entanglement can be interpreted in terms of entanglement between two quasi-particles, which keeps to contribute to $\Delta S^{(n)}_A$ and it approaches a constant:
\begin{equation} \label{see}
\Delta S^{(n \ge 1)}_A =\log{2}.
\end{equation}

As explained above, the late time value of $\Delta S^{(n)}_A$ for (\ref{exam}) comes from entanglement between an entanglement pair.
Then we assume that $\phi$ can be decomposed into the right and left moving modes ($\phi^L, \phi^R$), respectively:
\be
\begin{split}
\phi= \phi^{L\dagger}+\phi^{R\dagger} + \phi^{L}+\phi^{R},
\end{split}
\ee 
where they obey the following late time algebra:
\be\label{del}
\begin{split}
&\left[\phi^{M}, \phi^{N \dagger}\right]=\delta_{MN}, ~(N,M=R,L)\\
&\left[\phi^{M}, \phi^{N}\right]=0. \\
\end{split}
\ee
The ground state is decomposed into the ground states for the right and left moving modes:
\be \label{gdp}
\ket{0}=\ket{0}_R\otimes \ket{0}_L,
\ee
where 
\be
\phi^L\ket{0}_L=\phi^R\ket{0}_R=0.
\ee
In this picture, the excited state can be represented by
\be
\ket{\Psi_s}=\f{1}{\sqrt{2}}\left[\ket{\phi^L}_L\ket{0}_R +\ket{0}_L\ket{\phi^R}_R\right].
\ee
In the late time limit,  the right and left moving modes are included in $A$ and $B$, respectively in this case. Then the right and left moving modes can be identified with the physical degrees of freedom in $A$ and $B$, respectively. 
Therefore the effective density matrix $\rho^e_A$ is given by
\be \label{erd}
\rho^e_A =\Tr_B \ket{\Psi_s}\bra{\Psi_s} =\Tr_L\ket{\Psi_s}\bra{\Psi_s}=\f{1}{2}\left[\ket{\phi_R}\bra{\phi_R}+\ket{\phi_L}\bra{\phi_L}\right],
\ee
where $\ket{\phi^L}=\phi^{L\dagger}\ket{0}_L, \ket{\phi^R}=\phi^{R\dagger}\ket{0}_R$.
 (R\'enyi) entanglement entropy whose reduced density matrix is given by (\ref{erd}) is the same as (\ref{see}).

\subsection{$\rho^e_A$ Without Taking the Late Time Limit}
As in the previous subsection, the late time value of $\Delta S^{(n)}_A$ comes from the entanglement between quasi-particles. In other words, it can be given by $S_A^{(n)}$ whose reduced density matrix is given by the probability distribution of quasi-particles as follows. Here we assume that if a composite operator $:\phi^k:$ is inserted, each $\phi$ creates one quasi-particle. $k$ is an integer number.
\subsubsection{Reduced Density Matrix and Probability}
(\ref{erd}) shows that reduced density matrix can be thought as the probability distribution of the quasi-particle.
If we assume that a quasi-particle is created by $\phi(-t, -l, {\bf x})$, it is included in $A$ or $B$ with the probabilities $P_A(t)$ and $P_B(t)$ at $t=0$.
In this case, the particle should propagate spherically at the speed of light.
Then in the late time, it is included in $A$ or $B$ with $\lim_{t\rightarrow \infty}P_A(t)=\f{1}{2}$ and $\lim_{t\rightarrow \infty}P_B(t)=\f{1}{2}$.
They are the same as the components ($\bra{\phi^R}\rho_A\ket{\phi^R}$, $\bra{\phi^L}\rho_A\ket{\phi^L}$) of the effective reduced density matrix.  

Even if the late time limit is not taken, the effective density matrix $\rho^{e}_A(t)$ (the probability distribution of quasi-particles)  is assumed to be applicable.
The decomposition in (\ref{del}) is generalized as follows:
\be \label{adp}
\phi(-t, -l, {\bf x})= \phi^{L\dagger}(-t, -l, {\bf x})+\phi^{R\dagger} (-t, -l, {\bf x})+ \phi^{L}(-t, -l, {\bf x})+\phi^{R}(-t, -l, {\bf x}),
\ee
where they obey the following algebra:
\be \label{gal}
\begin{split}
&\left[\phi^{M}(-t, -l, {\bf x}), \phi^{N\dagger}(-t, -l, {\bf x})\right]=\delta_{MN}f^{M}(-t, -l, {\bf x }), ~(N,M=R,L)\\
&\left[\phi^{M}(-t, -l, {\bf x}), \phi^{N}(-t, -l, {\bf x})\right]=0. \\
\end{split}
\ee
Without taking the late time limit, the ground state is assumed to be decomposed in the same manner as in (\ref{gdp}).
However, the definition of the ground states for the left and right moving modes are generalized as follows: 
\be
\phi^L(-t, -l, {\bf x})\ket{0}_L=\phi^R(-t, -l, {\bf x})\ket{0}_R=0.
\ee
Here the norm of $\phi^{L\dagger}(-t, -l, {\bf x})\ket{0}_L,~ \phi^{R\dagger}(-t, -l, {\bf x})\ket{0}_R$ is given by
\be
\bra{0}_M\phi^M(-t, -l, {\bf x})\phi^{M \dagger}(-t, -l, {\bf x}) \ket{0}_M = f^{M} (-t, -l ,{\bf x}),
\ee 
where $M=L,R$. $\bra{0}_L\phi^L(-t, -l, {\bf x})\phi^{L \dagger}(-t, -l, {\bf x}) \ket{0}_L$ and $\bra{0}_R\phi^R(-t, -l, {\bf x})\phi^{R \dagger}(-t, -l, {\bf x}) \ket{0}_R$ correspond to the probabilities with which a quasi-particle is included in $B$ and $A$ respectively. Under the decomposition in (\ref{adp}), the state in (\ref{exam}) is represented by
\be
\ket{\Psi_s(-t)}=\mathcal{N}\left[\sqrt{f^L(-t, -l, {\bf x})}\ket{\phi^L(-t, -l , {\bf x})}_L\ket{0}_R +\sqrt{f^R(-t, -l, {\bf x})}\ket{0}_L\ket{\phi^R(-t, -l , {\bf x})}_R\right],
\ee
where  $\ket{\phi^M(-t, -l , {\bf x})}=\f{1}{\sqrt{f^M(-t, -l, {\bf x})}}\phi^{M\dagger}(-t, -l , {\bf x})\ket{0}_M$ and $\mathcal{N}^{-2}=f^L(-t, -l ,{\bf x})+f^R(-t, -l ,{\bf x})$.

If the effective reduced density matrix $\rho^e_A(t)$ is defined by
\be
\begin{split}
\rho^e_A(t) &:=\Tr_B \ket{\Psi_s(-t)} \bra {\Psi_s(-t)} \\
&=\Tr_L \ket{\Psi_s(-t)} \bra {\Psi_s(-t)} =P_1(t) \ket{0}_R \bra{0}_R+P_2(t)\ket{\phi^L(-t, -l ,{\bf x})}\bra{\phi^L(-t, -l ,{\bf x})},
\end{split}
\ee 
where $P_1(t) =\mathcal{N}^2f^L(-t, -l, {\bf x}), P_2(t) =\mathcal{N}^2f^R(-t, -l, {\bf x})$, then $\Delta S^{(n)}_A$ for $\rho_A^e(t)$ is given by
\be\label{sp}
\Delta S^{(n)}_A =\f{1}{1-n}\log{\Tr_R\left(\rho^e_A\right)^n} =\f{1}{1-n}\log{\left[(P_1(t))^n+(P_2(t))^n\right]}.
\ee

{\bf Diagrams}

\begin{figure}[htbp]
  \begin{center}
    \begin{tabular}{c}

      \begin{minipage}{0.5\hsize}
        \begin{center}
          \includegraphics[clip, width=6.5cm]{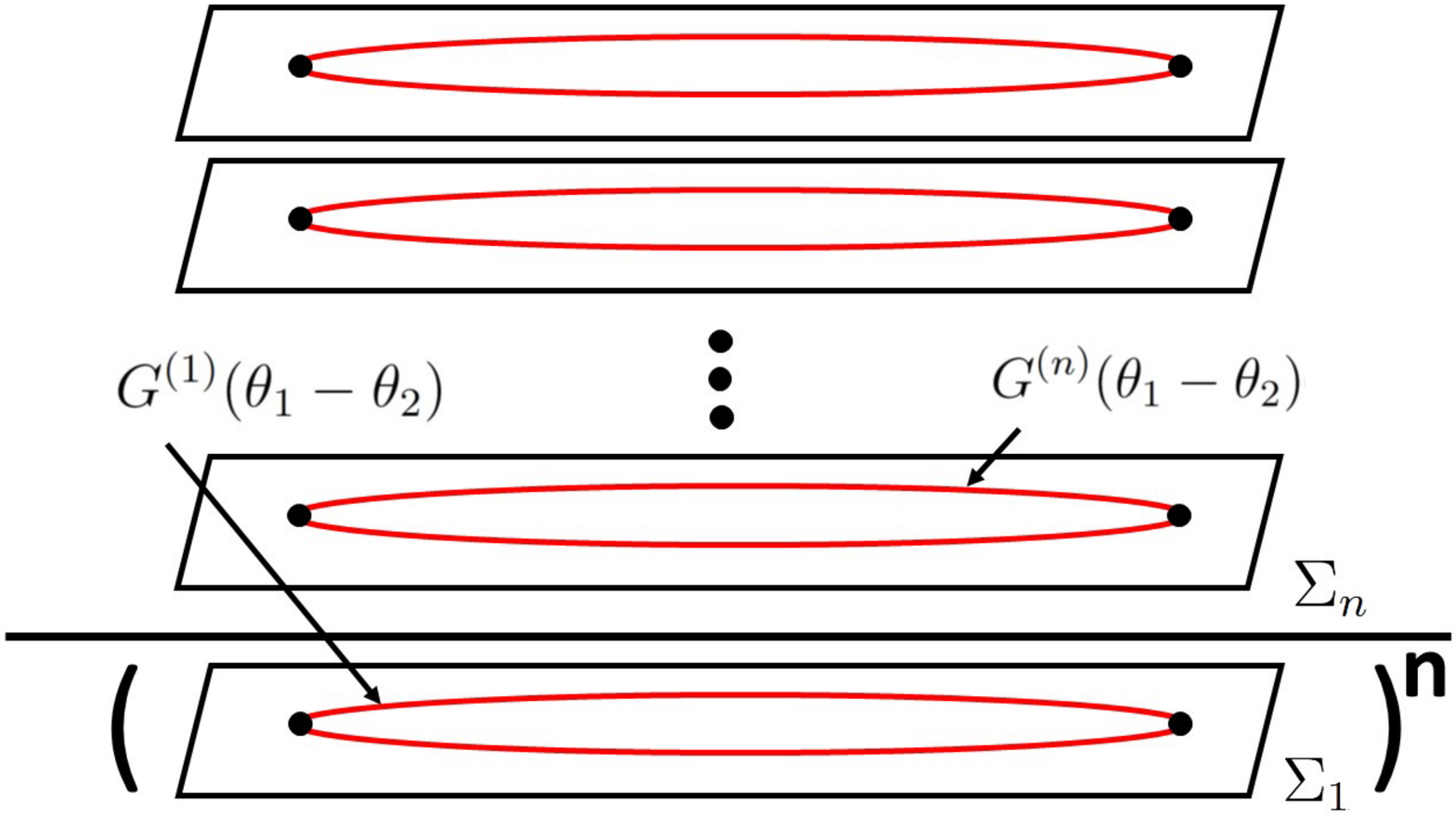}
          \hspace{1.6cm} \\
       (a)$\f{D^{(n)}_1}{(D_1)^n}$.
        \end{center}
      \end{minipage}

      \begin{minipage}{0.5\hsize}
        \begin{center}
          \includegraphics[clip, width=6.5cm]{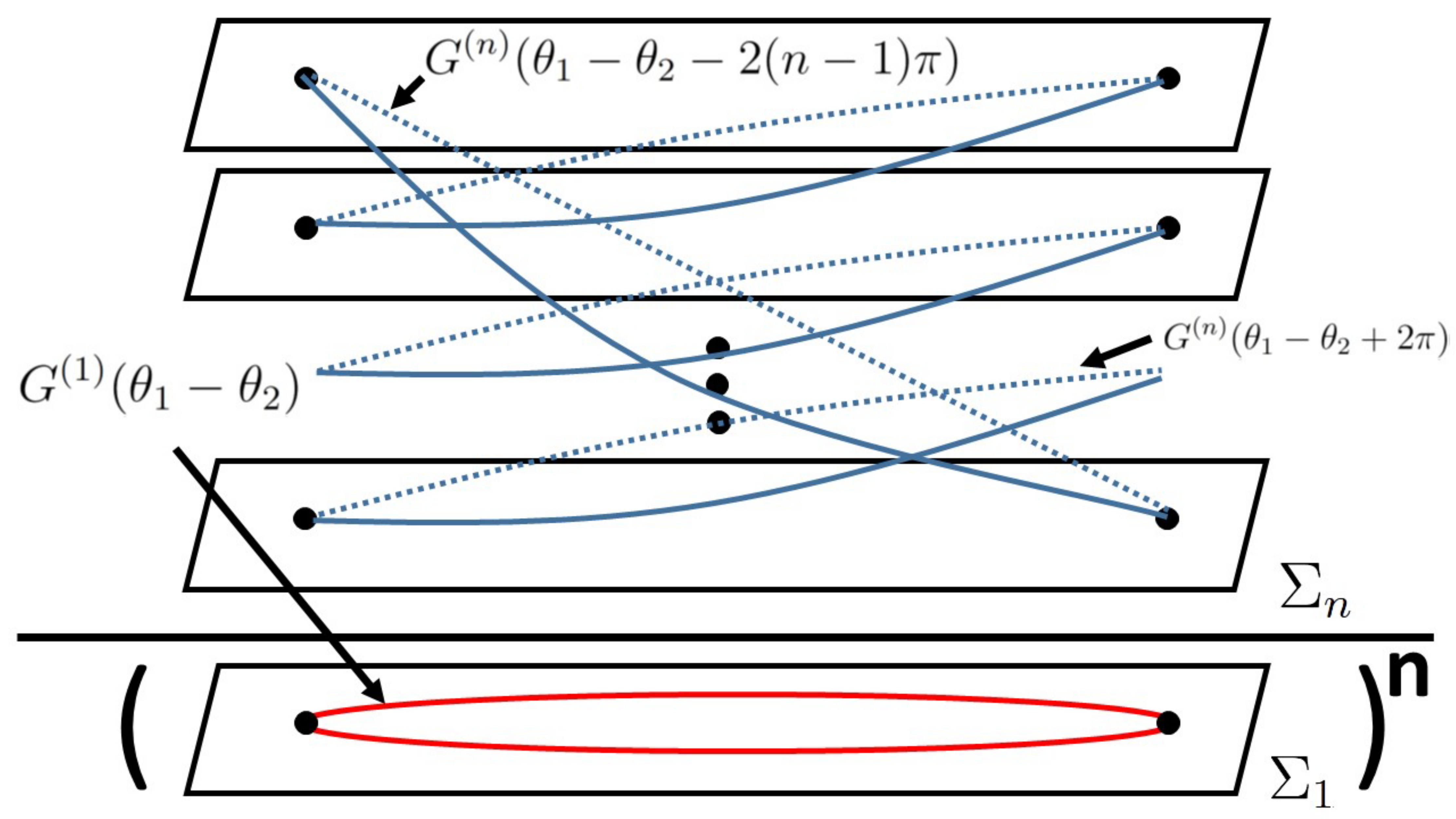}
          \hspace{1.6cm} \\
 (b)   $\f{D^{(n)}_2}{(D_1)^n}$.
        \end{center}
      \end{minipage}

    \end{tabular}
    \caption{The ratios of diagrams. The diagram constructed of $G^{(n)}(\theta_1-\theta_2)$ is called $D^{(n)}_1$. The diagram constructed of $G^{(n)}(\theta_1-\theta_2+2\pi)$ and $G^{(n)}(\theta_1-\theta_2-2(n-1)\pi)$ is $D^{(n)}_2$. The one constructed of $G^{(1)}(\theta_1-\theta_2)$ is $D^{(1)}_1$. (a) is the ratio of $D^{(n)}_1$ to $(D^{(1)}_1)^n$. (b) is the ratio of $D^{(n)}_2$ to $(D~{(1)}_1)^n$. }
    \label{fig:lena2}
  \end{center}
\end{figure}

In the limit $\epsilon \rightarrow 0$, $\Delta S^{(n)}_A$ in the replica method can be computed by a few diagrams. Green's functions used in the following are the leading orders in a small $\epsilon$ expansion.  
In $t \le l$, the diagram constructed of Green's function on the same sheet (Fig.\ref{fig:lena2} (a)) can contribute to $\Delta S^{(n)}_A$:
\be\label{sg1}
\Delta S^{(n)}_A=\f{1}{1-n} \log{\left[\f{\left(G^{(n)}(\theta_1-\theta_2)\right)^n}{\left(G^{(1)}(\theta_1-\theta_2)\right)^n}\right]},
\ee
where $G^{(n)}(\theta_1-\theta_2)$ and $G^{(1)}(\theta_1-\theta_2)$ are Green's functions on $\Sigma_n$ and $\Sigma_1$, respectively.
Green's functions for any $n$ have the property $G^{(n)}(\Theta)=G^{(n)}(-\Theta)$. 
The quasi-particle which is created by $\phi$ is included in $B$.
Then $P_2(t)$ has to vanish because $P_2(t)$ is the probability with which the particle is included in $A$. 
If (\ref{sp}) is identified with (\ref{sg1}), $P_1(t)$ is given by
\be \label{r1}
P_1(t)=\f{G^{(n)}(\theta_1-\theta_2)}{G^{(1)}(\theta_1-\theta_2)}.
\ee

In $t>l $ the other diagram (Fig.\ref{fig:lena2} (b)) constructed of $G^{(n)}(\theta_1-\theta_2+2\pi)$ and $G^{(n)}(\theta_1-\theta_2-2(n-1)\pi)$  can contribute to $\Delta S^{(n)}_A$, but $G^{(n)}(\theta_1-\theta_2+2\pi)=G^{(n)}(\theta_1-\theta_2-2(n-1)\pi)$.
Then $P_2(t)$ can be identified with the ratio of $G^{(n)}(\theta-\theta_2+2\pi)$ to $G^{(1)}(\theta-\theta_2)$:
\be \label{r2}
P_2(t)=\f{G^{(n)}(\theta_1-\theta_2+2\pi)}{G^{(1)}(\theta_1-\theta_2)}.
\ee
The ratio of $P_1(t)$ to $P_2(t)$ is given by
\be
\f{P_1(t)}{P_2(t)}=\f{f^L(-t, -l, {\bf x})}{f^R(-t, -l, {\bf x})}=\f{G^{(n)}(\theta_1-\theta_2)}{G^{(n)}(\theta_1-\theta_2+2\pi)}.
\ee
Then Green's functions can be chosen as  $f^{L, R}$ as follows \footnote{We do not claim that this choice is unique. There is an ambiguity of the overall factor of Green's functions. Here their factors are chosen so that (\ref{gal2}) in the late time limit satisfies (\ref{gal}).}:
\be \label{rl}
f^L(-t, -l, {\bf x})=32 \pi^2 \epsilon^2G^{(n)}(\theta_1-\theta_2),~ f^R(-t, -l, {\bf x})=32 \pi^2 \epsilon^2G^{(n)}(\theta_1-\theta_2+2\pi),
\ee
where $G^{(1)}(\theta-\theta_2)$ is given by the sum of Green's functions on $\Sigma_n$:
\be \label{id}
G^{(1)}(\theta-\theta_2)= G^{(n)}(\theta-\theta_2)+ G^{(n)}(\theta-\theta_2+2\pi).
\ee 
(\ref{rl}) satisfies (\ref{r1}) and (\ref{r2}). The sum of $P_i(t)$ is equal to $1$.
Then the algebra which quasi-particles obey is given by
\be \label{gal2}
\begin{split}
&\left[\phi^{L}(-t, -l, {\bf x}), \phi^{L\dagger}(-t, -l, {\bf x})\right]=32 \pi^2 \epsilon^2 G^{(n)}(\theta-\theta_2), \\
&\left[\phi^{R}(-t, -l, {\bf x}), \phi^{R\dagger}(-t, -l, {\bf x})\right]=32 \pi^2 \epsilon^2 G^{(n)}(\theta-\theta_2+2\pi), \\
&\left[\phi^{M}(-t, -l, {\bf x}), \phi^{N}(-t, -l, {\bf x})\right]=0.
\end{split}
\ee
(\ref{gal2}) shows  the commutation relation for $\phi^L$ ($\phi^R$) are given by the {Green's} function on the same sheet (Green's function on the different sheet) if $\phi$ is located in $B$\footnote{If the subregion $A$ is given by $x_1\le 0$, the commutation relation for $\phi^L$ ($\phi^R$) are given by $G^{(n)}(\theta-\theta_2+2\pi)$ ($G^{(n)}(\theta-\theta_2)$). }.  In the late time limit, the algebra in (\ref{gal2}) satisfies the late time algebra in (\ref{gal}). 
This relation between the commutation relation and Green's function holds in free Maxwell theory and they are summarized in appendix\footnote{ Although it is expected that the relation holds even in free fermionic theories, we did not check it.}.

Using the analytic continued Green's functions summarized in the appendix, (\ref{rl}) shows that $P_1(t)$ and $P_2(t)$ in $4$ dimensional massless free scalar theory is given by
\be \label{repp}
\begin{split}
&P_1(t)=\begin{cases}
1 & t\le l,\\
\f{t+l}{2t} &t>l, 
\end{cases} \\
&P_2(t)=\begin{cases}
0 & t\le l,\\
\f{t-l}{2t} &t>l.
\end{cases} 
\end{split}
\ee
{\bf A Simple Example}

Here we compute $\Delta S^{(n)}_A$ for a simple example by using the algebra in (\ref{gal2}). 
The given state is 
\be
\ket{\Psi}= \mathcal{N} :\phi^k:(-t, -l, {\bf x} )\ket{0}. 
\ee
where $\mathcal{N}$ is given by
\be
\mathcal{N}=\f{1}{k! (32 \pi^2 \epsilon^2)^k (G^{(n)}(\theta-\theta_2)+G^{(n)}(\theta-\theta_2+2\pi))^k}.
\ee
Its effective reduced density matrix is given by\footnote{${}_kC_l$ is a binomial coefficient defined by ${}_kC_l := \frac{(k)!}{l! (k-l)!}$.}
\be\label{efd}
\rho_A=\Tr_L \rho=\sum_{l=0}^k\f{_k C_l \left(G^{(n)}\left(\theta-\theta_2\right)\right)^{k-l}\left(G^{(n)}\left(\theta-\theta_2+2\pi\right)\right)^{l}}{\left(G^{(n)}(\theta-\theta_2)+G^{(n)}(\theta-\theta_2+2\pi)\right)^k}\ket{\left(\phi^{R}\right)^l}\bra{\left(\phi^{R}\right)^l}.
\ee
Then $S^{(n)}_A$ for this density matrix in (\ref{efd}) is given by
\be \label{St}
\begin{split}
&S^{(n)}_A=\f{1}{1-n}\log{\left[\sum_{l=0}^k\left(\f{_k C_l \left(G^{(n)}\left(\theta-\theta_2\right)\right)^{k-l}\left(G^{(n)}\left(\theta-\theta_2+2\pi\right)\right)^{l}}{\left(G^{(n)}(\theta-\theta_2)+G^{(n)}(\theta-\theta_2+2\pi)\right)^k}\right)^n\right]} \\
&= \f{1}{1-n}\log{\left[\sum_{l=0}^k~_k C_l(P_1(t))^l(P_2(t))^{k-l}\right]}.
\end{split}
\ee
where we use the identity in (\ref{id}).
The entropy in the late time limit is given by
\be \label{Sl}
 S^{(n)}_A=\f{1}{1-n}\log{\left[\f{\left(_k C_l\right)^n}{2^{k n}}\right]}.
\ee

These results in (\ref{St} ) and (\ref{Sl}) are consistent with the results in the replica method\cite{MN1, MN2}.

\section{Particle Propagating Model}


Here we consider a toy model where we can explain what $\Delta S^{(n)}_A$ measures in free field theories.
For simplicity, the theory we consider is $4$ dimensional free massless scalar field theory.
In this toy model, a local operator creates quasi-particles and their probability changes with respect to time. 
As an example, consider the case that a local operator $\phi$ is acting on the ground state at $(t, x_1, x_2, x_3)=(0, -l ,0, 0)$.
It creates a quasi-particle at $(t, x_1, x_2, x_3)=(0, -l ,0, 0)$.
The particle propagates spherically at the speed of light without any interactions. 
The total system is divided into $A$ and $B$. $A$ ($B$) is given by $x_1 \ge 0$ $(x_1 < 0)$. 
 
\begin{figure}[h]
\centering
	\centering
	\includegraphics[width=60mm]{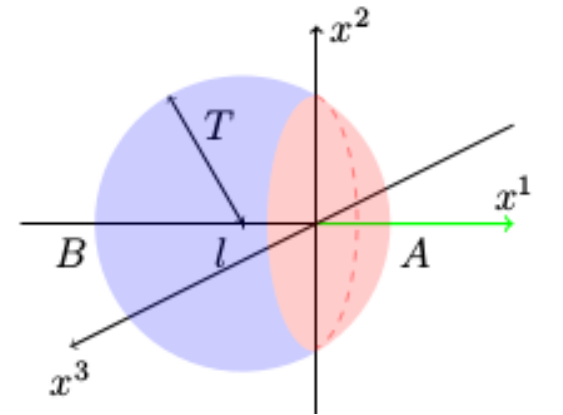}
	\caption{Free particle propagation}
 \label{fig:fpprop}
\end{figure}

At $t=T (\le l)$, the particle is necessarily included in $B$.
The probability $P_A(T)$, with which the quasi-particle is included in $A$, should vanish.
On the other hand,  $P_B(T)$, with which the quasi-particle is included in $B$, is equal to $1$.
Then the probability distribution $\rho$ is defined by
\be
\rho:=P_B(T) \ket{0, 1}\bra{0, 1}+P_A(T) \ket{1, 0}\bra{1, 0},
\ee
where $\ket{l,n}$ is the state where $l$ ($n$) particles are included in $A$ ($B$) with $P_A(T)$ ($P_B(T)$).
We defined (R\'enyi) entropy $S^{(n\ge 1)}$ for $\rho$ by
\be
\begin{split} \label{stoy}
S^{(n)}:= \begin{cases}
\f{1}{1-n}\log{\Tr\left(\rho^n\right)} &n \ge 2, \\
-\Tr \rho \log{\rho} & n=1.
\end{cases}
\end{split}
\ee
 At $t=T$,  the particle stays somewhere on the sphere whose radius is $T$ as in Fig. \ref{fig:fpprop}.
$S^{(n)}$ for $\rho$ vanishes for $T \leq l$.
For $T>l$, a part of the spherical surface of the quasi-particles' propagation is included in region $A$.
The area $\mathcal{S}_A$ of the part of surface which is included in the region $A$ is
\be \label{area}
\mathcal{S}_A(T)=\int ^{\alpha}_0 d\theta 2 \pi T^2 \sin{\theta} d\theta =  2 \pi T^2 (1-\f{l}{T})=2\pi T (T-l),
\ee   
where $\cos{\alpha}=\f{l}{T}$.
$P_A(T)$ ($P_B(T)$) is given by the ratio of $\mathcal{S}_A(T)$ $(\mathcal{S}_B(T))$ to the area $\mathcal{S}_{all}(T)$ of the surface with the radius $T$ :
\be \label{tdop}
P_A(T) =\f{\mathcal{S}_A(T)}{\mathcal{S}_{all}(T)}, ~P_B(T) =\f{\mathcal{S}_B(T)}{\mathcal{S}_{all}(T)}.
\ee 
Thus the probabilities with which the particle is included in region $A$ and $B$ in $T>l$ are  
\be \label{model}
\begin{split}
&P_A(T) = \f{2\pi T (T-l)}{4\pi T^2} = \f{(T-l)}{2T},\\
& P_B (T)=\f{(T+l)}{2T}.
\end{split}
\ee

The probabilities in (\ref{model}) are consistent with the ones in (\ref{repp}). $S^{(n)}$ for $\rho $ is consistent with $\Delta S^{(n)}_A$ in the replica method.

Thus, with this toy model, $\Delta S^{(n)}_A$ for $\phi$ in $4d$ can be reproduced.
Here we implicitly assume that particles propagate isotropically. For the particle with spin, it is expected that the weight is changed from $1$ to $W(t, x)$, which depends on the particle's spin.
For the particle with spin, the integration in (\ref{area}) might be changed to
\be
\mathcal{S}'_A(T)=\int_{A } dV  W(T, \theta).
\ee 
where the integration at $t=T$ is performed for the part of spherical surface, which is included in $A$ as in Fig.\ref{fig:fpprop}. The definition of probabilities in (\ref{tdop}) changes to 
\be 
P_A(T) =\f{\mathcal{S}'_A(T)}{\mathcal{S}'_{all}(T)}, ~P_B(T) =\f{\mathcal{S}'_B(T)}{\mathcal{S}'_{all}(T)}.
\ee 

Let's compute $\Delta S^{(n)}_A$ and $\Delta I_{A,B}$ with our model. 

\subsection{Example.1: $\mathcal{O}=:\phi^k:( 0, -l,  {\bf x})$}
Here the local operator $:\phi^k:$ which is located at  $(t, x_1, x_2, x_3)=(0, -l ,0, 0)$ acts on the ground state. The following assumption is taken. {\it The $k$ same-kind particles are created at the point where the local composite operator is inserted since it is constructed of only $\phi$}. Then in $T>l$, the $l$ particles and $k-l$ particles are included in $A$ and $B$ with the probability $_k C_l \left( P_A(T) \right)^{k-l}\left( P_B(T) \right)^{l}$. Thus the probability distribution $\rho$ is given by
\be \label{mpd}
\rho=\sum_{l=0}^{k} ~_k C_l \left( P_A(T) \right)^{k-l}\left( P_B(T) \right)^{l}\ket{l, k-l}\bra{l, k-l},
\ee
where $\ket{l, k-l}$ is the state where $l$ and $k-l$ particles are included in $A$ and $B$, respectively. $S^{(n)}$ for (\ref{mpd}) is consistent with (\ref{St}).  

\subsection{Example.2 : $\mathcal{O}=\phi(-T,-L, {\bf x}_1) \phi(-t, -l, {\bf x}_2) $}
The given state is 
\be
\ket{\Psi} = \mathcal{N} \phi(-T,-L, {\bf x}_1) \phi(-t, -l, {\bf x}_2) \ket{0},
\ee
where $l, L, t , T>0$. Here we assume that {\it a particle created by $\phi(-T, -L, {\bf x}_1)$ is a different kind particle from the one created by $\phi(-t,-l, {\bf x}_2)$}. The distribution $\rho$ at $t=0$ is defined by
\be \label{sepr}
\begin{split}
\rho&= \sum_{a,b,c,d} P_{a, c} \tilde{P}_{b, d} \ket{a,b; c, d} \bra{a,b; c, d} \\
&=P_{1, 0}\tilde{P}_{1,0} \ket{1,1,0,0}\bra{1,1,0,0}+P_{0, 1}\tilde{P}_{1, 0} \ket{0,1,1,0}\bra{0,1,1,0}\\
&+P_{0, 1}\tilde{P}_{0,1} \ket{1,0,0,1}\bra{1,0,0,1}+P_{0, 1}\tilde{P}_{0, 1} \ket{0,0,1,1}\bra{0,0,1,1},
\end{split}
\ee
where $\ket{a,b; c, d}$ is the state where $a$ (b) and $c$ (d) particles created by $\phi(-T, -L, {\bf x}_1) $ ($ \phi(-t, -l, {\bf x}_2)$) are included in $A$ and $B$, respectively. 
Each probability at $t=0$ is given by
\be
\begin{split}
&P_{1,0}=\begin{cases}
0 &T<L,\\
\f{(T-L)}{2T} & T\ge L,
\end{cases},~~ 
P_{0,1}=\begin{cases}
1 &T<L ,\\
\f{(T+L)}{2T} & T\ge L,
\end{cases}\\
&\tilde{P}_{1,0}=\begin{cases}
0 &t<l, \\
\f{(t'-l)}{2(t)} & t\ge l,
\end{cases},~~
\tilde{P}_{0,1}=\begin{cases}
1 &t<l, \\
\f{(t+l)}{2(t)} & t\ge l,
\end{cases}\\
\end{split}
\ee

$S^{(n)}$ for (\ref{sepr}) is given by
\be
\begin{split}
&S^{(n>1)}=\f{1}{1-n}\log{\left[\left(P_{1, 0}\tilde{P}_{1,0} \right)^n+\left(P_{ 0, 1}\tilde{P}_{1,0} \right)^n+\left(P_{1,0}\tilde{P}_{0,1} \right)^n+\left(P_{0, 1}\tilde{P}_{0, 1} \right)^n\right]}, \\
&S=S^{(n=1)}=-\left(P_{1, 0}\tilde{P}_{1,0} \right)\log{\left(P_{1, 0}\tilde{P}_{1,0} \right)}-\left(P_{ 0, 1}\tilde{P}_{1,0} \right)\log{\left(P_{ 0, 1}\tilde{P}_{1,0} \right)} \\
&-\left(P_{1,0}\tilde{P}_{0,1} \right)\log{\left(P_{1,0}\tilde{P}_{0,1} \right)}-\left(P_{0, 1}\tilde{P}_{0, 1} \right)\log{\left(P_{0, 1}\tilde{P}_{0, 1} \right)}.
\end{split}
\ee

In the late time limit, they are given by $S^{(n\ge 1)}=\log{4}$ which is consistent with the result in \cite{MN2}.

\subsection{Example.3: A Finite Interval}
\begin{figure}[h]
\centering
  \begin{center}
   \includegraphics[width=100mm]{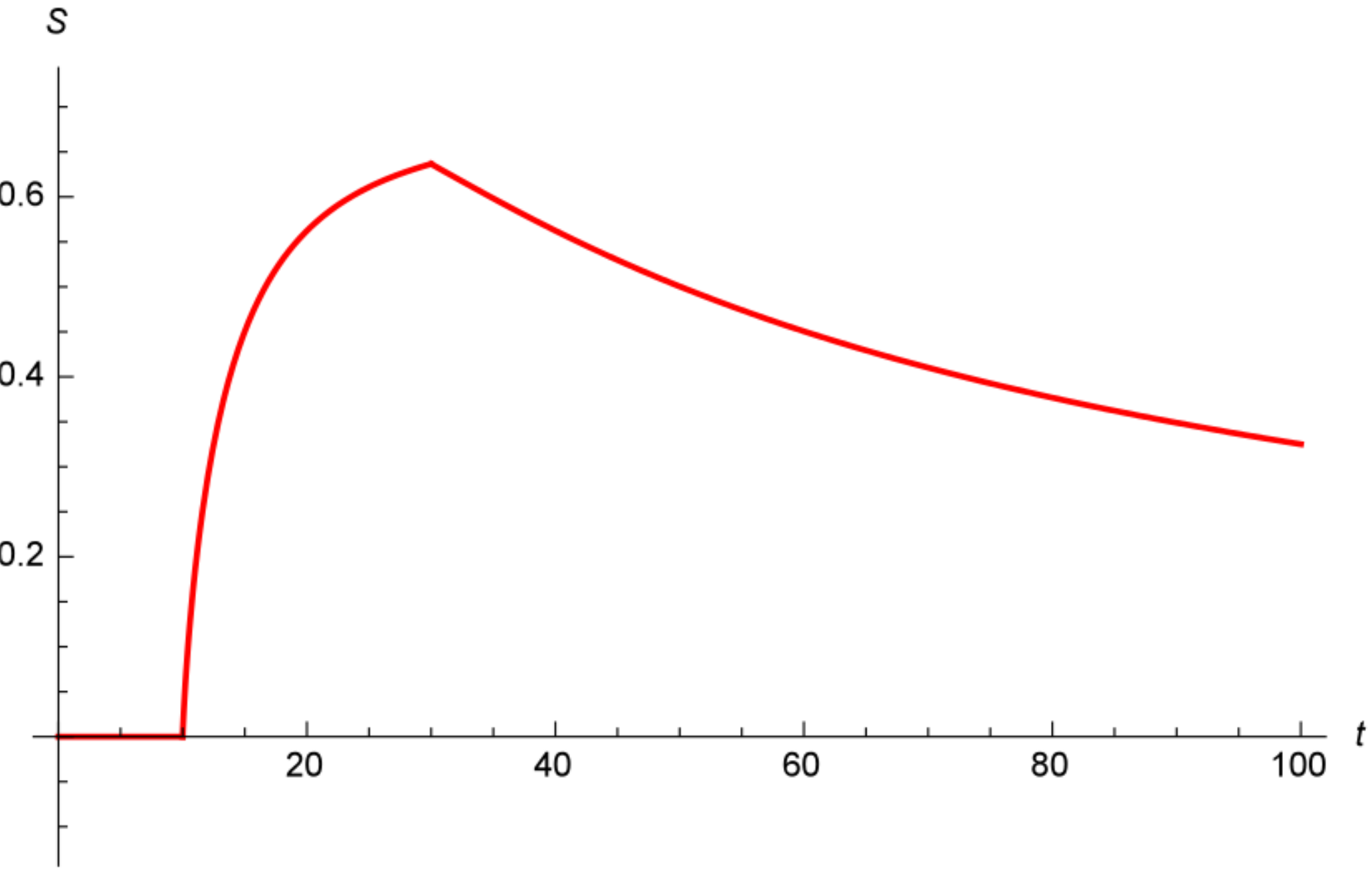}
  \end{center}
 \caption{The Plot of $S$.The parallel axis is $t$. The vertical axes is $S$. $(l, L)=(10,30)$. }
 \label{F_I}
\end{figure}

The local operator $\phi$ is located at $(t, x_1, {\bf x})=(0, 0, {\bf 0})$ and the given subsystem is $0<l \le x_1<L$. $S^{(n\ge1)}$ is measured at $t$. A quasi-particle is included in $A$ ( $B$) with the probability $P_A(t)$ $(P_B(t))$. They are given by
\be
\begin{split}
&P_A(t) =\begin{cases}
0& 0 \le t<l, \\
\f{t-l}{2t} & 0<l \le t<L, \\
\f{L-l}{2t}& L \le t,  
\end{cases} , ~~
P_B(t) =\begin{cases}
1& 0 \le t<l, \\
\f{t+l}{2t} & 0<l \le t<L, \\
\f{L+l}{2t} & L \le t,  
\end{cases}
\end{split}
\ee 
whose $S^{(n\ge 1)}$ are given by
\be
\begin{split}
& S^{(n\ge 1)}=\begin{cases}
0 & 0<t<l, \\
 \f{1}{1-n}\log{\left[(\f{t-l}{2t})^n+(\f{t+l}{2t})^n\right]}& 0<l\le t<L, \\
\f{1}{1-n}\log{\left[(1-\f{L-l}{2t})^n+(\f{L-l}{2t})^n\right]}& L\le t, \\
\end{cases} \\
& S=\begin{cases}
0 & 0<t<l, \\
-\left(\f{t-l}{2t}\right)\log{\left(\f{t-l}{2t}\right)}-\left(\f{t+l}{2t}\right)\log{\left(\f{t+l}{2t}\right)}& 0<l\le t<L, , \\
-\left(1-\f{L-l}{2t}\right)\log{\left(1-\f{L-l}{2t}\right)}-\left(\f{L-l}{2t}\right)\log{\left(\f{L-l}{2t}\right)} & L\le t. \\
\end{cases}
\end{split}
\ee
The plot of $ S$ shows that $S$ increases after $t=l$ and decreases after $t=L$ (Fig.\ref{F_I}).  $P_A(t)$ increases in $l<t<L$ but it decreases before it approaches $\f{1}{2}$. Therefore, it does not approach $S$ for the maximally entangled state and vanishes at the late time.

\subsection{Example.4: Infinite Subsystems}
Here the given subsystem is infinite but its shape is more complicated than the one discussed previously.
The following subsystems are considered:
\be
\begin{split}
&A_1=\{x_1\ge l, x_2\ge 0\},~~ A_2=\{x_1\ge l, x_2 \ge 0, x_3 \ge 0\}.
\end{split}
\ee
The subsystem $B$ is defined by the remnant of the total space. 
The probability distribution in this case is defined by 
\be
\rho_i:=P_B(T) \ket{0, 1}\bra{0, 1}+P_{A_i}(T) \ket{1, 0}\bra{1, 0}, (i=1 \sim 2)
\ee
where the probabilities are given by
\be
\begin{split}
&A_1: P_{A_1}=\f{1}{4}\left(1-\f{l}{t}\right), ~~P_{B}=\f{1}{4}\left(3+\f{l}{t}\right), \\
&A_2: P_{A_2}=\f{1}{8}\left(1-\f{l}{t}\right), ~~P_B=\f{1}{8}\left(7+\f{l}{t}\right).
\end{split}
\ee
 $\ket{l, k-l}$ is the state where $l$ and $k-l$ particles are included in $A_i$ and $B$ respectively.  
Their entropies in $t \le l$ vanish. They in $t>l$ are given by
\be
\begin{split}
& S^{(n)}(\rho_1)= 
\begin{cases}
\f{1}{1-n}\log{\left[\left(\f{1}{4}\left(1-\f{l}{t}\right)\right)^n+\left(\f{1}{4}\left(3+\f{l}{t}\right)\right)^n\right]} & n \ge 2, \\
-\f{1}{4}\left(1-\f{l}{t}\right)\log{\left[ \f{1}{4}\left(1-\f{l}{t}\right)\right]}- \f{1}{4}\left(3+\f{l}{t}\right)\log{\left[\f{1}{4}\left(3+\f{l}{t}\right)\right]} &n=1, \\
\end{cases} \\
&S^{(n)}(\rho_2)= 
\begin{cases}
\f{1}{1-n}\log{\left[\left(\f{1}{8}\left(1-\f{l}{t}\right)\right)^n+\left(\f{1}{8}\left(7+\f{l}{t}\right)\right)^n\right]} & n \ge 2, \\
-\f{1}{8}\left(1-\f{l}{t}\right)\log{\left[ \f{1}{8}\left(1-\f{l}{t}\right)\right]}- \f{1}{8}\left(7+\f{l}{t}\right)\log{\left[\f{1}{8}\left(7+\f{l}{t}\right)\right]} &n=1. \\
\end{cases} \\
\end{split}
\ee
Since the particle created by $\phi$ can stay at $A_{1,2}$ or $B$ in the late time limit, the entropies in the limit are finite:
\be
\begin{split}
& S^{(n)}(\rho_1)= 
\begin{cases}
\f{1}{1-n}\log{\left[\left(\f{1}{4}\right)^n+\left(\f{3}{4}\right)^n\right]} & n \ge 2, \\
-\f{1}{4}\log{\left[ \f{1}{4}\right]}- \f{3}{4}\log{\left[\f{3}{4}\right]} &n=1, \\
\end{cases} \\
& S^{(n)}(\rho_2)= 
\begin{cases}
\f{1}{1-n}\log{\left[\left(\f{1}{8}\right)^n+\left(\f{7}{8}\right)^n\right]} & n \ge 2, \\
-\f{1}{8}\log{\left[ \f{1}{8}\right]}- \f{7}{8}\log{\left[\f{7}{8}\right]} &n=1. \\
\end{cases} \\
\end{split}
\ee
$S(\rho_{i=1,2})$ are smaller than the entropy for an EPR state.
\subsection{Mutual Information}
The mutual information $I(A,B)$ measures the correlation between $A$ and $B$\cite{MI1, MI2, MI3, MI4, MI5, MI6}.
Here the excess of mutual information $\Delta I_{A, B}$ is defined by subtracting the mutual information  for the ground state $I^G_{A, B}$ from  for the locally excited state $I^{EX}_{A, B}$: 
\be
\Delta I_{A,B}=I^{EX}_{A,B}- I^G _{A,B}=\Delta S_A+\Delta S_B-\Delta S_{A \cup B},
\ee
where $\Delta S_{A \cup B}$ is the excess of mutual entanglement entropy for $A$ or $B$. In our toy model, $\Delta S^{(n)}_A$ for a locally excited state is evaluated by computing $S^{(n)}$ for a probability distribution $\rho$.
\subsubsection{$\Delta I_{A, B}$ between a finite interval and infinite interval}
The total space is divided into the three regions $A, B$ and $C$. They are given by
\be
\begin{split}
A: {0<l_A  \le x_1 },~~ B:{-L_B\le x_1 \le -l_B<0}, ~~ C: \text{the remnant of the total space}.
\end{split}
\ee 
The local operator $\phi$ is located at $(t, x_1, {\bf x})=(0, 0, {\bf 0})$. We compute $\Delta I_{A, B}$ in order to measures the time evolution of the correlation between the subregion $A$ and $B$ at $t$.  
The excess of the mutual information $\Delta I(A,B)$ is given by
\be
\begin{split}
&\Delta I_{A,B}=\Delta S_A+\Delta S_B-\Delta S_{A \cup B},
\end{split}
\ee
As explained earlier, $\Delta S_M$ is evaluated by $S(\rho_M)$. Thus, $\Delta I_{A,B}$ is evaluated by $\mathcal{I}_{A, B}$ which is defined by

\be
\begin{split}
&\mathcal{I}_{A,B}= S(\rho_A)+ S(\rho_B)- S(\rho_{A \cup B}),
\end{split}
\ee
where $S(\rho_M)$ are given by
\be
\begin{split}
&S(\rho_A)=-P_A(t)\log{P_A(t)}-P_{B \cup C}(t)\log{P_{B \cup C}(t)}, \\
&S(\rho_B)=-P_B(t)\log{P_B(t)}-P_{A \cup C}(t)\log{P_{A \cup C}(t)}, \\
&S(\rho_{A \cup B})=-P_{A\cup B}(t)\log{P_{A\cup B}(t)}-P_{C}(t)\log{P_{C}(t)}, \\
\end{split}
\ee
\subsubsection*{$\bullet~ 0< l_A < l_B < L_B$}
Here the parameters satisfy the following relation:
\be
0< l_A < l_B < L_B.
\ee

Since the particle created by $\phi$ stays at $-l_B < x_1 <l_A$ in $0\le t < l_A$, 
$P_C=1$ and $\mathcal{I}_{A,B}$ vanishes. In $l_A \le t < l_B$, it can be included in $B$. The probabilities are given by 
\be
\begin{split}
&P_A(t)=\f{t-l_A}{2t},~~ P_{A \cup C}=\f{t+l_A}{2t}, \\
&P_B(t)=0, ~~ P_{A \cup C}(t)=1, \\
&P_{A \cup B}(t)=\f{t-l_A}{2t},~~ P_C(t)=\f{t+l_A}{2t}.\\
\end{split}
\ee
Then $\mathcal{I}_{A,B}$ vanishes because $S(\rho_A)$ cancels with $S(\rho_{A \cup B})$.
It is expected that the correlation disappears because the particle is included  in only $A$.

The particle can stay in $A$ and $B$ in $0<l_B\le t \le L_B$. The probabilities are given by
\be \label{pc11}
\begin{split}
&P_A(t)=\left(\f{t-l_A}{2t}\right), ~~ P_{B \cup C}(t)=\left(\f{t+l_A}{2t}\right), \\
&P_B(t)=\left(\f{t-l_B}{2t}\right), ~~P_{A \cup C}(t)=\left(\f{t+l_B}{2t}\right), \\
&P_{A \cup B}(t)= \left(\f{2t-l_A-l_B}{2t}\right), ~~P_{C}(t)= \left(\f{l_A+l_B}{2t}\right).
\end{split}
\ee
The correlation between $A$ and $B$ increases because the particle can stay in both $A$ and $B$. 
\begin{figure}[htbp]
 \begin{minipage}{0.5\hsize}
  \begin{center}
   \includegraphics[width=60mm]{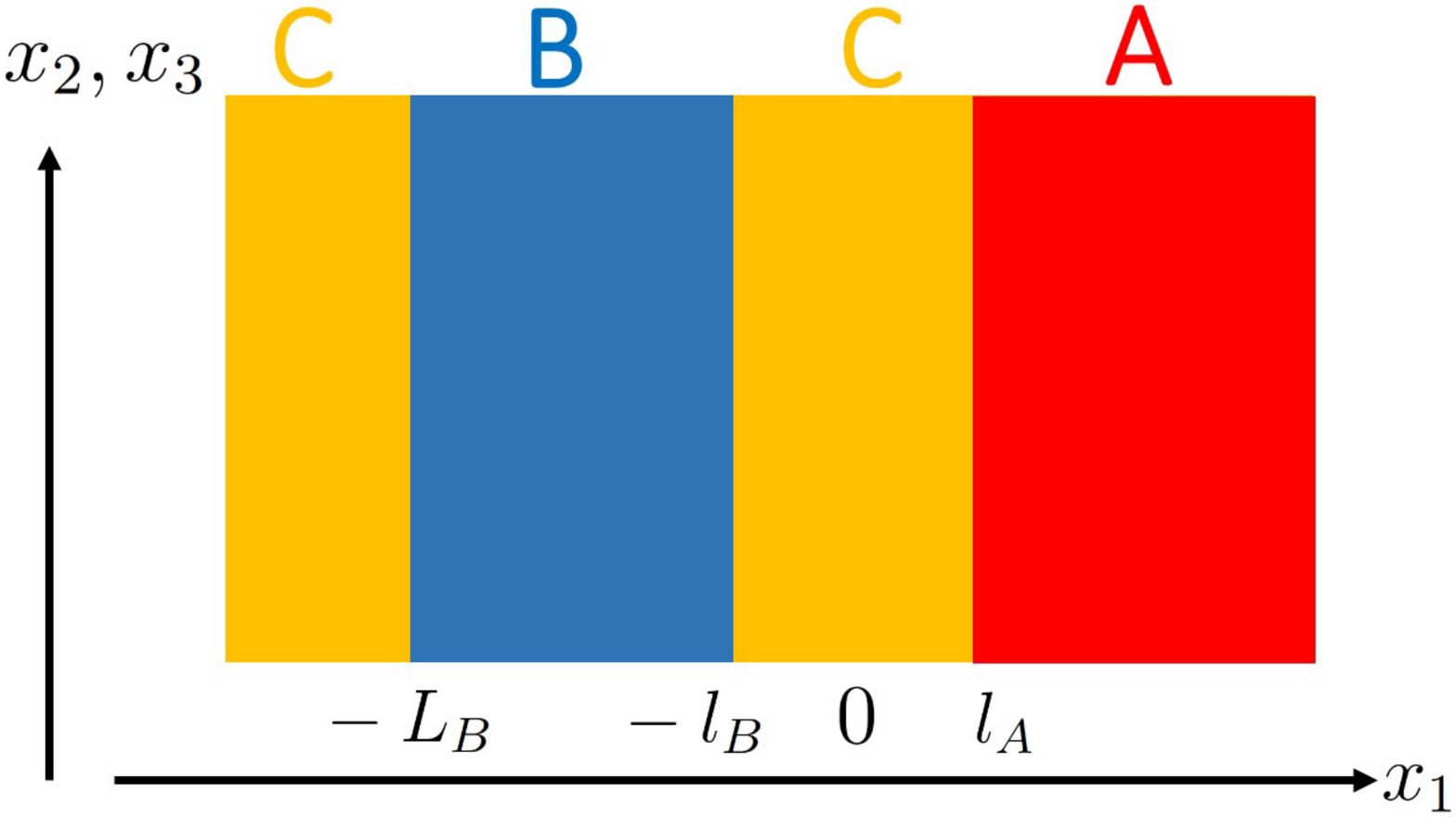}
  \end{center}
  \caption{A schematic explanation of the subsystems $A$, $B$ and $C$. $A$ is an infinite strip. $B$ is a finite strip.}
  \label{setinfi}
 \end{minipage}
\hspace{5pt}
 \begin{minipage}{0.5\hsize}
  \begin{center}
	\includegraphics[width=60mm]{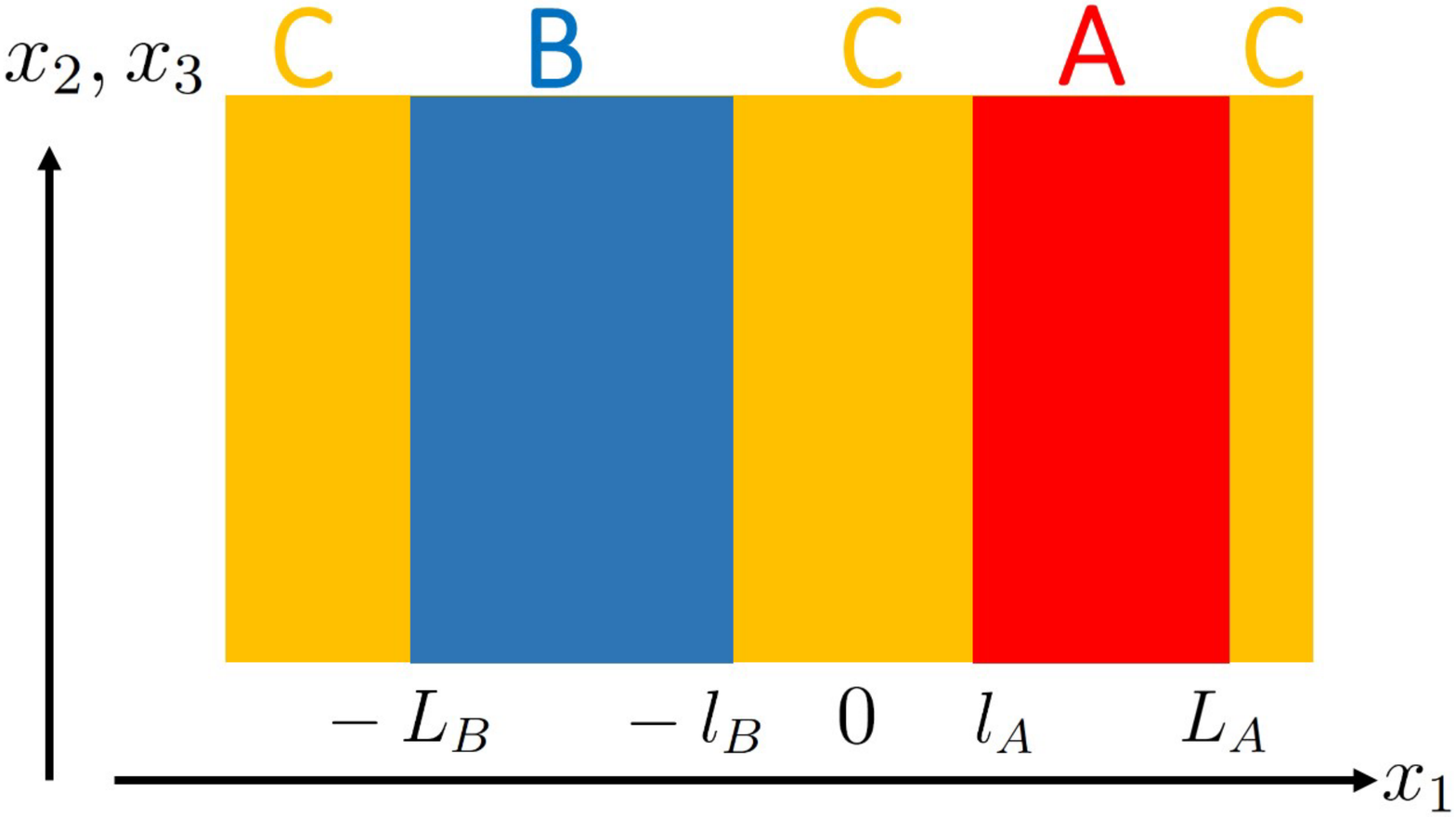}
  \end{center}
  \caption{A schematic explanation of the subsystems $A$, $B$ and $C$. $A$ and $B$ are  finite strips. }
  \label{setfifi}
 \end{minipage}
\end{figure}
\hspace{5pt}

The probabilities in $L_B \ge t$ are given by
\be \label{pc12}
\begin{split}
&P_A(t)=\left(\f{t-l_A}{2t}\right), ~~ P_{B \cup C}(t)=\left(\f{t+l_A}{2t}\right), \\
&P_B(t)=\left(\f{L_B-l_B}{2t}\right), ~~P_{A \cup C}(t)=\left(\f{2t-L_B+l_B}{2t}\right), \\
&P_{A \cup B}(t)= \left(\f{t+L_B-l_A-l_B}{2t}\right), ~~P_{C}(t)=  \left(\f{t-L_B+l_A+l_B}{2t}\right).
\end{split}
\ee

$P_B(t)$ decreases because the particle tends to come out of $B$ in this region. In the late time limit, the particle is outside $B$. Then $\mathcal{I}_{A,B}$ vanishes. The time evolution of $\mathcal{I}_{A,B}$ is plotted in Fig.\ref{infi}. 
\subsubsection*{$\bullet~ 0< l_B< l_A < L_B$}
Here the parameters considered obey that
\be
0< l_B< l_A < L_B.
\ee

\begin{figure}[htbp]
 \begin{minipage}{0.5\hsize}
  \begin{center}
   \includegraphics[width=70mm]{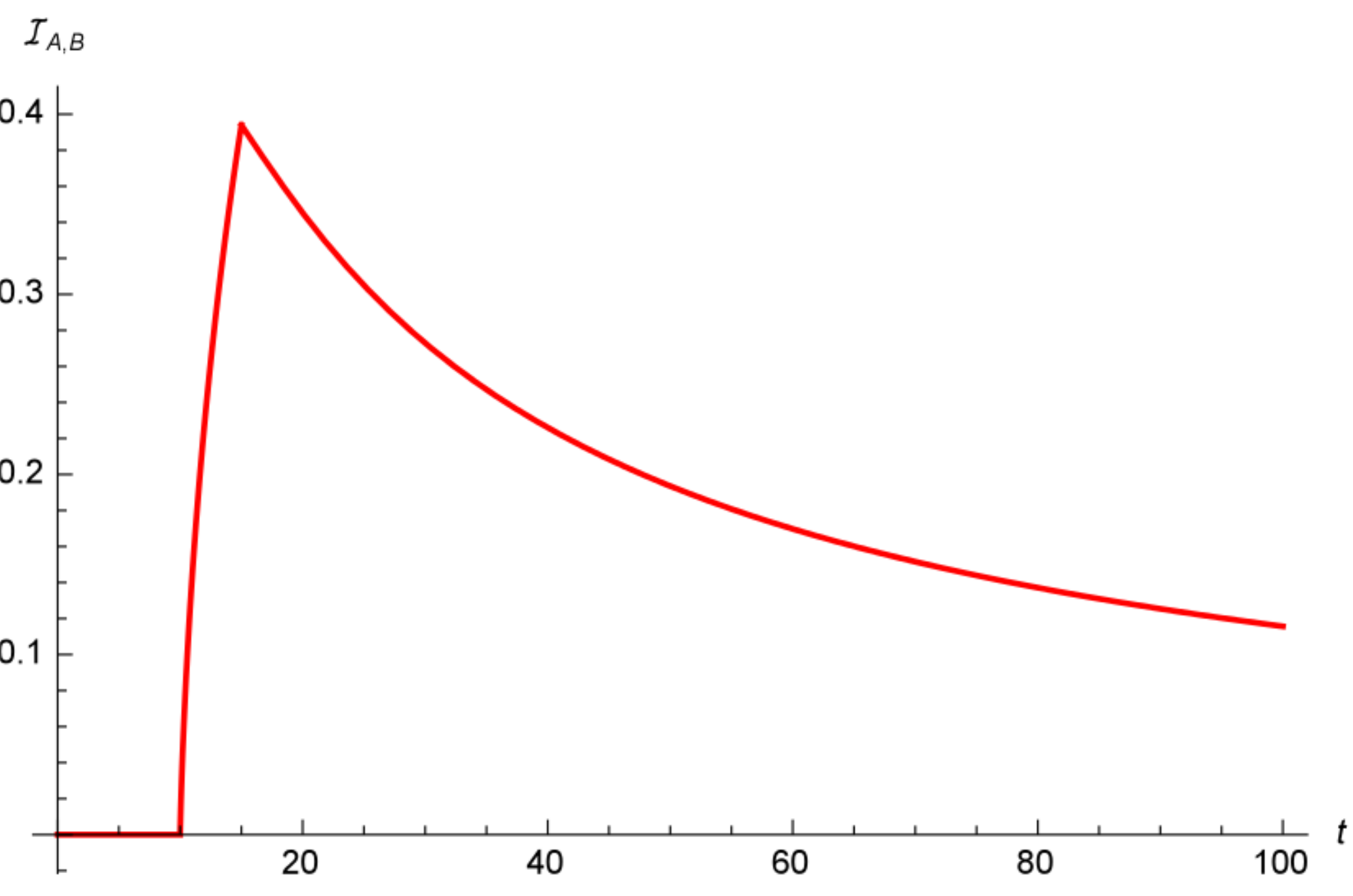}
  \end{center}
 \caption{The Plot of $\mathcal{I}_{A, B}$ .The horizontal axis is $t$. The vertical axis is $\mathcal{I}_{A, B}$ . $(l_A, l_B, L_B)=(5,10,15)$. }
\label{infi}
 \end{minipage}
\hspace{5pt}
 \begin{minipage}{0.5\hsize}
    \begin{center}
   \includegraphics[width=70mm]{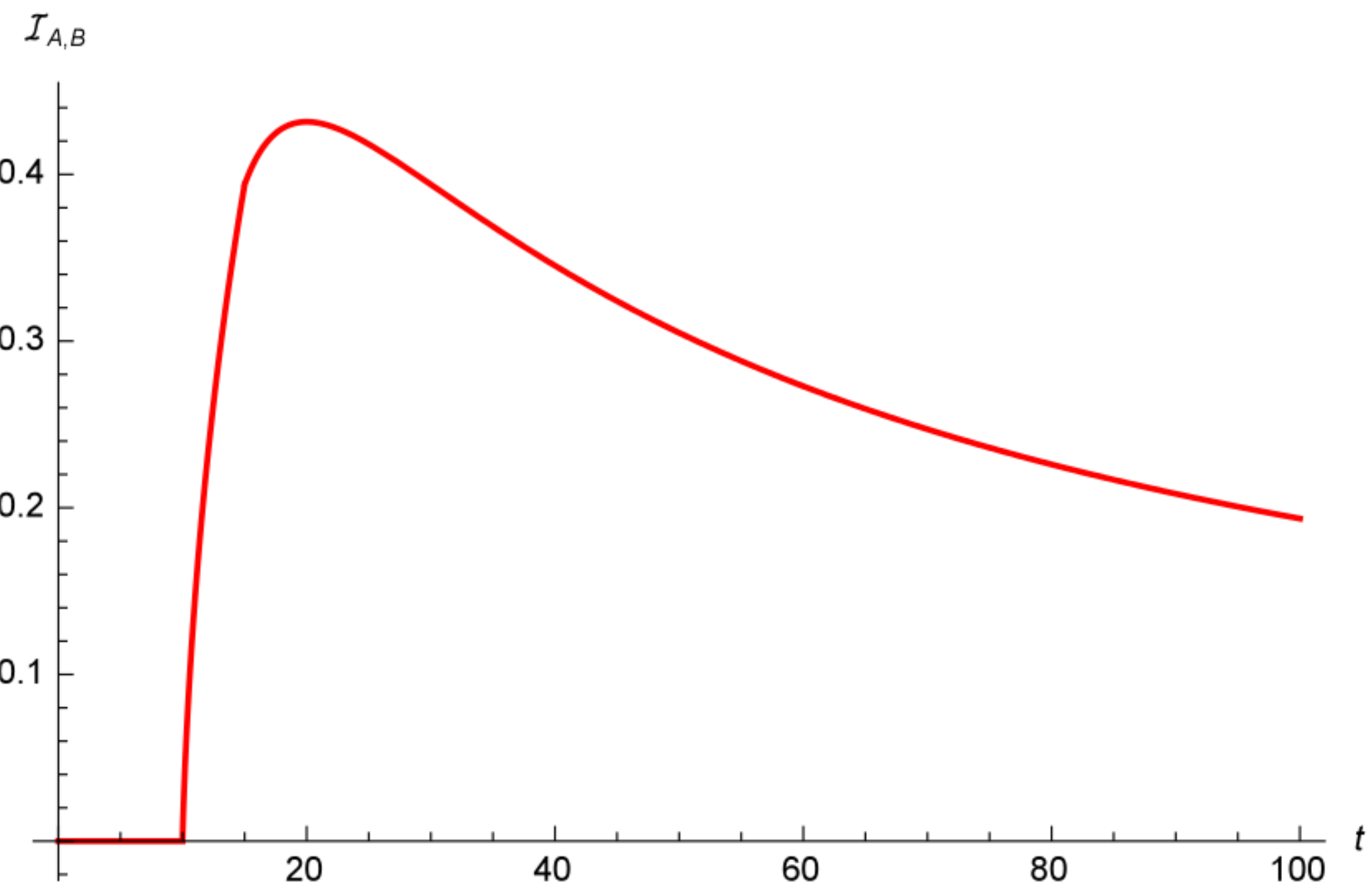}
  \end{center}
 \caption{The Plot of $\mathcal{I}_{A, B}$ .The horizontal axis is $t$. The vertical axis is $\mathcal{I}_{A, B}$ . $(l_A, l_B, L_B)=(10,5,15)$. }
\label{infi2}
 \end{minipage}
\hspace{5pt}
\end{figure}

In $0 < t \le l_A$, $\mathcal{I}_{A, B}$ vanishes. 
In $l_A < t \le L_B$, the probabilities are the same as (\ref{pc11}).
Those in $t>L_B$ are the same as (\ref{pc12}). 
In $t <L_B$ the time evolution of $\mathcal{I}_{A,B}$ does not depend on whether $l_A$ is greater or smaller than $l_B$.
It depends on the relation between $l_A$ and $l_B$ only if $t>L_B$.    
The time evolution of $\mathcal{I}_{A,B}$ is plotted in Fig.\ref{infi2}. 
\subsubsection{$\Delta I(A,B)$ between two finite intervals} 
Here we evaluate $\Delta I_{A,B}$ for the two finite intervals by computing $\mathcal{I}_{A, B}$.
The given subsystems are
\be
\begin{split}
A: {0<l_A  \le x_1 \le L_A},~~ B:{-L_B\le x_1 \le -l_B<0}, ~~ C: \text{the remnant of the total space}.
\end{split}
\ee 

\subsubsection*{$\bullet l_A<l_B<L_A<L_B$}
Here we assume that $l_A < l_B <L_A<L_B$.
$\Delta I(A, B)$ vanishes because the particle is necessarily included in $C$ in $t \le l_A$.
Since the particle is necessarily outside $B$, the probabilities in $l_A \le t < l_B$ are given by 
\be
\begin{split}
&P_A(t)=\left(\f{t-l_A}{2t}\right), ~~ P_{B \cup C}(t)=\left(\f{t+l_A}{2t}\right), \\
&P_B(t)=0, ~~P_{A \cup C}(t)=1, \\
&P_{A \cup B}(t)= \left(\f{t-l_A}{2t}\right), ~~P_{C}(t)=  \left(\f{t+l_A}{2t}\right).
\end{split}
\ee
 It is expected that $\mathcal{I}_{A,B}$ vanishes because the particle can stay in $A$ but can not stay in $B$.  

It can stay in both $A$ and $B$ in $l_B \le t < L_A$ and probabilities are given by 
 \be \label{p1}
\begin{split}
&P_A(t)=\left(\f{t-l_A}{2t}\right), ~~ P_{B \cup C}(t)=\left(\f{t+l_A}{2t}\right), \\
&P_B(t)=\left(\f{t-l_B}{2t}\right), ~~P_{A \cup C}(t)=\left(\f{t+l_B}{2t}\right), \\
&P_{A \cup B}(t)= \left(\f{2t-(l_A+l_B)}{2t}\right), ~~P_{C}(t)=  \left(\f{l_A+l_B}{2t}\right).
\end{split}
\ee 
$\mathcal{I}_{A,B}$ increases in this region.

In $L_A \le t < L_B$, the particle can come out of $A$. The probabilities are given by
 \be\label{p2}
\begin{split}
&P_A(t)=\left(\f{L_A-l_A}{2t}\right), ~~ P_{B \cup C}(t)=\left(\f{2t-L_A+l_A}{2t}\right), \\
&P_B(t)=\left(\f{t-l_B}{2t}\right), ~~P_{A \cup C}(t)=\left(\f{t+l_B}{2t}\right), \\
&P_{A \cup B}(t)= \left(\f{t+L_A-(l_A+l_B)}{2t}\right), ~~P_{C}(t)=  \left(\f{t-L_A+(l_A+l_B)}{2t}\right).
\end{split}
\ee 
In $0<L_B<t$, it comes out of $A$ and $B$. They are given by
\be\label{p3}
\begin{split}
&P_A(t)=\left(\f{L_A-l_A}{2t}\right), ~~ P_{B \cup C}(t)=\left(\f{2t-L_A+l_A}{2t}\right), \\
&P_B(t)=\left(\f{L_B-l_B}{2t}\right), ~~P_{A \cup C}(t)=\left(\f{2t-(L_B-l_B)}{2t}\right), \\
&P_{A \cup B}(t)= \left(\f{L_A+L_B-l_B-l_A}{2t}\right), ~~P_{C}(t)=  \left(\f{2t-L_B-L_A+(l_A+l_B)}{2t}\right).
\end{split}
\ee 
$\mathcal{I}_{A,B}$ decreases in this region. If we assume that the correlation between $A$ and $B$ comes from probabilities in $A$ and $B$, this behavior is reasonable.
$\mathcal{I}_{A,B}$ eventually vanishes. 
Its plot is shown in Fig.\ref{fifi}.

\subsubsection*{$l_A<l_B<L_B<L_A$}
Here we assume that $l_A<l_B<L_B<L_A$. Before $t=l_B$, $\mathcal{I}_{A,B}$ vanishes. The probabilities $P_i$ in $l_B \le t < L_B$ is the same as (\ref{p1}). $P_i$ in $L_B \le t < L_A$ is 
\be
\begin{split}
&P_A(t)=\left(\f{t-l_A}{2t}\right), ~~ P_{B \cup C}(t)=\left(\f{t+l_A}{2t}\right), \\
&P_B(t)=\left(\f {L_B-l_B}{2t}\right), ~~P_{A \cup C}(t)=\left(\f{2t-(L_B-l_B)}{2t}\right), \\
&P_{A \cup B}(t)= \left(\f{t+L_B-(l_A+l_B)}{2t}\right), ~~P_{C}(t)=  \left(\f{t-L_B+l_A+l_B}{2t}\right).
\end{split}
\ee 
Those in $L_A \le t$ are the same as (\ref{p3}). 

\begin{figure}[htbp]
 
\begin{minipage}{0.5\hsize}
  \begin{center}
   \includegraphics[width=70mm]{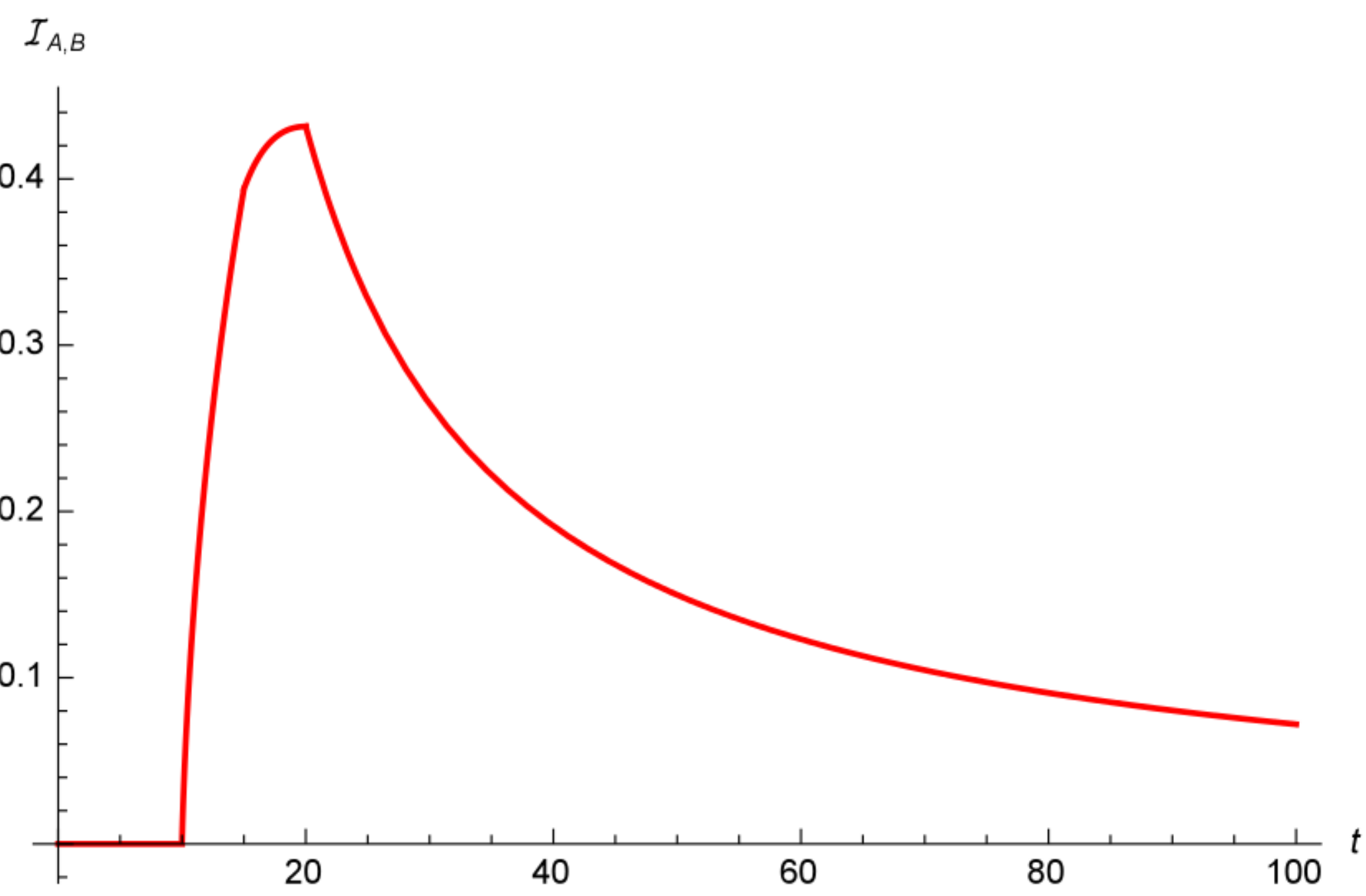}
  \end{center}
 \caption{The Plot of $\mathcal{I}_{A,B}$.The horizontal axis is $t$. The vertical axes is $\mathcal{I}_{A,B}$. $(l_A, l_B, L_A, L_B)=(5, 10, 15, 20)$. }
\label{fifi}
 \end{minipage}
\hspace{5pt}
\begin{minipage}{0.5\hsize}
    \begin{center}
   \includegraphics[width=70mm]{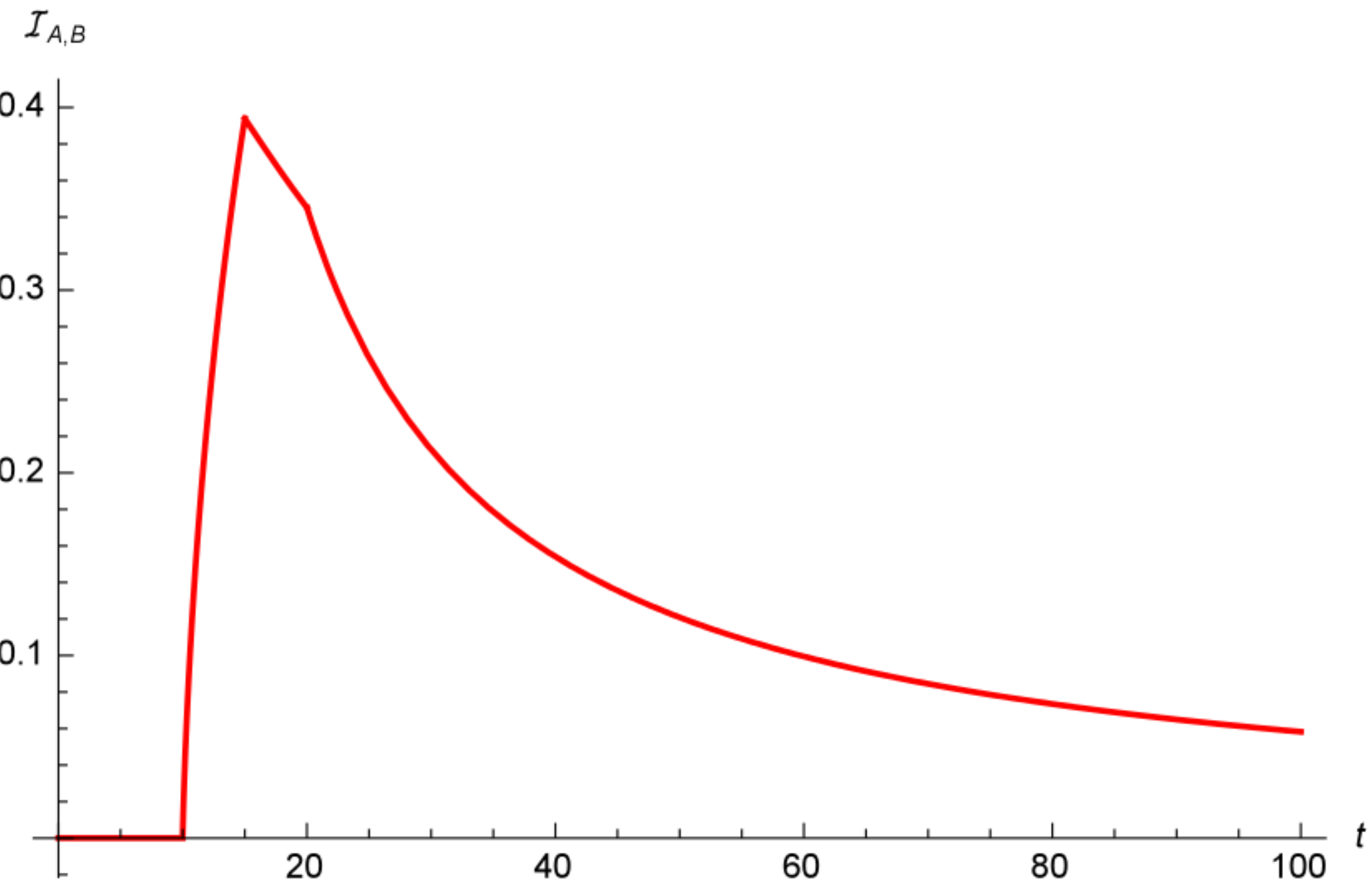}
  \end{center}
 \caption{The Plot of $\mathcal{I}_{A,B}$.The horizontal axis is $t$. The vertical axes is $\mathcal{I}_{A,B}$.  $(l_A, l_B, L_A, L_B)=(5, 10, 20, 15)$.}
\label{fifi2}
 \end{minipage}
\hspace{5pt}
\end{figure}

 The time evolution of $\mathcal{I}_{A, B}$  is plotted in Fig.\ref{fifi2}.
$\mathcal{I}_{A,B}$ eventually vanishes. 

The results in this section seem to show that the nontrivial time evolution of $\mathcal{I}_{A,B}$ appears if the particle can stay in $A$ and $B$ with the probabilities $P_A(t)$ and $P_B(t)$. 
 
Here we assume $4$ dimensional massless free scalar theory. We expect that our toy model can be generalized to higher dimensional cases and to other theories.  
\section{Summary and Discussion}
In this paper, we have shown that $\Delta S^{(n)}_A$ for locally excited states can be quantitatively interpreted in terms of quasi-particles even if the late time limit is not taken.
 $\Delta S^{(n)}_A$ is given by  (R\'enyi) entanglement entropy whose reduced density matrix is given by probability distribution of the quasi-particles.
The commutation relations which the quasi-particles obey are related with Green's functions. 

We have proposed a toy model and checked that it can reproduce the results in the replica method in $4$ dimensional free massless scalar field theory. The assumptions taken are:

\begin{itemize}
	\item[1] A local operator which is not a composite operator  creates a particle which propagates spherically without any interactions. For example, $\phi$ creates a quasi-particle which propagates spherically at the speed of light.
	\item[2] The composite operator constructed of only one species of operator creates one kind of quasi-particle. For example, $:\phi^k:$ creates $k$ quasi-particles of the same kind.
	\item[3]  If a operator is inserted at a different point from the point where another is located, it creates a different kind of particle.
	\item[4] $\Delta S^{(n)}_A$ can be evaluated by computing the entropies in (\ref{stoy}) for the probability distribution $\rho$ of the particles created by local operators.
\end{itemize}

In this paper, we have studied what $\Delta S^{(n)}_A$ measures in a simple system.
In this case, it is the distribution of quasi-particles.
The authors in \cite{MM1} proposed a model which explains dynamics of entanglement in the global quenches.
In that model, it is explained by the collection motion of quasi-particles.
We expect that there is a relation between our model and theirs.
It is one of the interesting future problems. 

In the global quenches, if the massive theory with the mass $m$ is suddenly changed to CFT, there is a scale $\xi =1/m$. We assumed that entanglement entropy is measured at $t$. If $t > \xi$, the quasi-particle picture can be applied, even though $\xi$ corresponds to $\epsilon$ in our case, $\epsilon$ can be taken $0$. In a holographic theory, the limit $\epsilon \rightarrow 0$ can not be taken. Therefore, it is interesting to study what $\epsilon  (\xi) $ is physically. It is also interesting to study whether the limit can be taken in a weakly interacting theory which is not integrable.

A generalization of our toy model to higher dimensional theories and other theories is not difficult and it is interesting. 
It is important that one computes $\Delta S^{(n)}_A$ and $\Delta I_{A,B}$ in the replica method and check they are consistent with the results by our toy model.

Our model can not explain the result in the minimal model \cite{TKO1} and holographic theory \cite{MN5} quantitatively. We expect that there is some mechanism which explains their results qualitatively. It will show what is the fundamental object which carries quantum entanglement. We hope that the object will clarify the fundamental mechanism beyond the $AdS/CFT$ correspondence more. 

\section*{Acknowledgement} 
We thank Tadashi Takayanagi, Pawel Caputa and Sumit Das for the useful discussions. 
NW thanks also Tadakatsu Sakai for helpful discussions.
\newpage
\appendix
\section{Commutations and Propagators}
Here we summarize the commutation relation for the quasi-particles and propagators in $4$ and $6$ dimensional free Maxwell theories and $4$ dimensional free massless theory.
\subsection{$4$ dimensional free massless scalar theory}
\subsubsection*{Propagators}
Analytic continued Green's functions for $n=1$ are given by 
\be\label{sgn1}
\begin{split}
G^{(1)}(\theta_1-\theta_2)=\f{1}{16\pi^2\epsilon^2}.
\end{split}
\ee
The functions for any $n$ in $0<t<l$ are the same as (\ref{sgn1}).
Those for any $n$ in $0<l\le t$ are given by
\be
\begin{split}
&G^{(n)}(\theta_1-\theta_2)=G^{(n)}(\theta_2-\theta_1)=\f{t+l}{32\pi^2t\epsilon^2}, \\
&G^{(n)}(\theta_1-\theta_2+2\pi)=G^{(n)}(\theta_2-\theta_1-2\pi)\\
&=G^{(n)}(\theta_1-\theta_2+2(n-1)\pi)=G^{(n)}(\theta_2-\theta_1-2(n-1)\pi)=\f{t-l}{32\pi^2t\epsilon^2},
\end{split}
\ee
where there is an identity:
\be
G^{(1)}(\theta_1-\theta_2)=G^{(n)}(\theta_1-\theta_2)+G^{(n)}(\theta_1-\theta_2+2\pi).
\ee

\subsubsection*{The Commutation Relation}
The commutation relation is given in the main text.
\subsection{$4$ dimensional Maxwell Theory}
\subsubsection*{Propagators}
The electric and magnetic fields $E_i, B_i$ are defined by
\be 
E_i= F_{0 i}, ~B_1=-F_{23}, ~B_2=F_{13}, ~B_3=-F_{12}.
\ee 
The analytic continued Green's functions are defined by
\be
\begin{split}
&\left\langle E_1(\theta)E_1(\theta')\right\rangle=F_{E1E1}(\theta-\theta'), \\
& \left\langle E_2(\theta)E_2(\theta')\right\rangle=\left\langle E_3(\theta)E_3(\theta')\right\rangle=F_{E2E2}(\theta-\theta'), \\
&\left\langle B_1(\theta)B_1(\theta')\right\rangle=F_{B1B1}(\theta-\theta'), \\
& \left\langle B_2(\theta)B_2(\theta')\right\rangle=\left\langle B_3(\theta)B_3(\theta')\right\rangle=F_{B2B2}(\theta-\theta'), \\
& \left\langle E_2(\theta)B_3(\theta')\right\rangle=F_{E2B3}(\theta-\theta'), \\
& \left\langle B_3(\theta)E_2(\theta')\right\rangle=F_{B3E2}(\theta-\theta'), \\
& \left\langle E_3(\theta)B_2(\theta')\right\rangle=F_{E3B2}(\theta-\theta'), \\
& \left\langle B_2(\theta)E_3(\theta')\right\rangle=F_{B2E3}(\theta-\theta'), \\
\end{split}
\ee
If the limit $\epsilon \rightarrow 0$ is taken, the leading term of them for $n=1$  are given by
\be \label{dpge}
\begin{split}
&F^{(1)}_{E1E1}(\theta_1-\theta_2) \sim \frac{1}{16 \pi ^2 \epsilon ^4}, \\
&F^{(1)}_{E2E2}(\theta_1-\theta_2)\sim \frac{1}{16 \pi ^2 \epsilon ^4}, \\
&F^{(1)}_{B1B1}(\theta_1-\theta_2)\sim \frac{1}{16 \pi ^2 \epsilon ^4}, \\
&F^{(1)}_{B2B2}(\theta_1-\theta_2)\sim \frac{1}{16 \pi ^2 \epsilon ^4}.
\end{split}
\ee

The propagators for $n \ge 2$ in $0<t \le l$ are the same as in (\ref{dpge}).

The propagators for arbitrary $n$ in $0<l \le t$ are given by
\be \label{dpgnl}
\begin{split}
&F^{(n)}_{E1E1}(\theta_1-\theta_2)=F^{(n)}_{E1E1}(\theta_2-\theta_1) =-\frac{(l-2 t) (l+t)^2}{64 \pi ^2 t^3 \epsilon ^4}, \\
&F^{(n)}_{E2E2}(\theta_1-\theta_2)=F^{(n)}_{E2E2}(\theta_2-\theta_1)= \frac{l^3+3 l t^2+4 t^3}{128 \pi ^2 t^3 \epsilon ^4}, \\
&F^{(n)}_{B1B1}(\theta_1-\theta_2)=F^{(n)}_{B1B1}(\theta_2-\theta_1)= -\frac{(l-2 t) (l+t)^2}{64 \pi ^2 t^3 \epsilon ^4}, \\
&F^{(n)}_{B2B2}(\theta_1-\theta_2)=F^{(n)}_{B2B2}(\theta_2-\theta_1)= \frac{l^3+3 l t^2+4 t^3}{128 \pi ^2 t^3 \epsilon ^4},\\
&F^{(n)}_{E2B3}(\theta_1-\theta_2)=F^{(n)}_{E2B3}(\theta_2-\theta_1) = \frac{3 (t-l) (l+t)}{128 \pi ^2 t^2 \epsilon ^4},\\
&F^{(n)}_{B3E2}(\theta_1-\theta_2)=F^{(n)}_{B3E2}(\theta_2-\theta_1) = \frac{3 (t-l) (l+t)}{128 \pi ^2 t^2 \epsilon ^4},\\
&F^{(n)}_{E3B2}(\theta_1-\theta_2)=F^{(n)}_{E3B2}(\theta_2-\theta_1) = \frac{3 (l-t) (l+t)}{128 \pi ^2 t^2 \epsilon ^4},\\
&F^{(n)}_{B2E3}(\theta_1-\theta_2)=F^{(n}_{B2E3}(\theta_2-\theta_1) = \frac{3 (l-t) (l+t)}{128 \pi ^2 t^2 \epsilon ^4},\\
\end{split}
\ee
\be \label{dpgnl2}
\begin{split}
&F^{(n)}_{E1E1}(\theta_1-\theta_2+2\pi)=F^{(n)}_{E1E1}(\theta_2-\theta_1-2\pi) \\
&=F^{(n)}_{E1E1}(\theta_1-\theta_2-2(n-1)\pi)=F^{(n)}_{E1E1}(\theta_2-\theta_1+2(n-1)\pi)= \frac{(l-t)^2 (l+2 t)}{64 \pi ^2 t^3 \epsilon ^4}, \\
&F^{(n)}_{E2E2}(\theta_1-\theta_2+2\pi)=F^{(n)}_{E2E2}(\theta_2-\theta_1-2\pi) \\
&=F^{(n)}_{E2E2}(\theta_1-\theta_2-2(n-1)\pi)=F^{(n)}_{E2E2}(\theta_2-\theta_1+2(n-1)\pi)=-\frac{l^3+3 l t^2-4 t^3}{128 \pi ^2 t^3 \epsilon ^4}, \\
&F^{(n)}_{B1B1}(\theta_1-\theta_2+2\pi)=F^{(n)}_{B1B1}(\theta_2-\theta_1-2\pi) \\
&=F^{(n)}_{B1B1}(\theta_1-\theta_2-2(n-1)\pi)=F^{(n)}_{B1B1}(\theta_2-\theta_1+2(n-1)\pi)= \frac{(l-t)^2 (l+2 t)}{64 \pi ^2 t^3 \epsilon ^4},\\
&F^{(n)}_{B2B2}(\theta_1-\theta_2+2\pi)=F^{(n)}_{B2B2}(\theta_2-\theta_1-2\pi)\\
&=F^{(n)}_{B2B2}(\theta_1-\theta_2-2(n-1)\pi)=F^{(n)}_{B2B2}(\theta_2-\theta_1+2(n-1)\pi)=-\frac{l^3+3 l t^2-4 t^3}{128 \pi ^2 t^3 \epsilon ^4}, \\
&F^{(n)}_{E2B3}(\theta_1-\theta_2+2\pi)=F^{(n)}_{E2B3}(\theta_2-\theta_1-2\pi) \\
&=F^{(n)}_{E2B3}(\theta_1-\theta_2-2(n-1)\pi)=F^{(n)}_{E2B3}(\theta_2-\theta_1+2(n-1)\pi)=\frac{3 (l-t) (l+t)}{128 \pi ^2 t^2 \epsilon ^4} \\ 
&F^{(n)}_{B3E2}(\theta_1-\theta_2+2\pi)=F^{(n)}_{B3E2}(\theta_2-\theta_1-2\pi)\\
&=F^{(n)}_{B3E2}(\theta_1-\theta_2-2(n-1)\pi)=F^{(n)}_{B3E2}(\theta_2-\theta_1+2(n-1)\pi)= \frac{3 (l-t) (l+t)}{128 \pi ^2 t^2 \epsilon ^4} ,\\
&F^{(n)}_{E3B2}(\theta_1-\theta_2+2\pi)=F^{(n)}_{E3B2}(\theta_2-\theta_1-2\pi) \\
&=F^{(n)}_{E3B2}(\theta_1-\theta_2-2(n-1)\pi)=F^{(n)}_{E3B2}(\theta_2-\theta_1+2(n-1)\pi)= \frac{3 (t-l) (l+t)}{128 \pi ^2 t^2 \epsilon ^4}, \\
&F^{(n)}_{B2E3}(\theta_1-\theta_2+2\pi)=F^{(n)}_{B2E3}(\theta_2-\theta_1-2\pi)\\
&=F^{(n)}_{B2E3}(\theta_1-\theta_2-2(n-1)\pi)=F^{(n)}_{B2E3}(\theta_2-\theta_1+2(n-1)\pi)=\frac{3 (t-l) (l+t)}{128 \pi ^2 t^2 \epsilon ^4}.\\
\end{split}
\ee
The contribution of the other propagators is much smaller than those in (\ref{dpgnl}) and  (\ref{dpgnl2}). 
They satisfy the following identities: 
\be
\begin{split}
&F^{(1)}_{E_iE_i}(\theta_1-\theta_2)=F^{(n)}_{E_iE_i}(\theta_1-\theta_2)+F^{(n)}_{E_iE_i}(\theta_1-\theta_2+2\pi), \\
&F^{(1)}_{B_iB_i}(\theta_1-\theta_2)=F^{(n)}_{B_iB_i}(\theta_1-\theta_2)+F^{(n)}_{B_iB_i}(\theta_1-\theta_2+2\pi), \\
&F^{(n)}_{E_2B_3}(\theta_1-\theta_2)+F^{(n)}_{E_2B_3}(\theta_1-\theta_2+2\pi)=0, \\
&F^{(n)}_{B_2E_3}(\theta_1-\theta_2)+F^{(n)}_{B_2E_3}(\theta_1-\theta_2+2\pi)=0, \\
\end{split}
\ee
\subsubsection{The commutation relation}
The electric and magnetic operators $E_i(-t, -l, {\bf x})$, $B_i(-t, -l, {\bf x})$ are decomposed into the left and right moving modes as follows:
\be \label{qps}
\begin{split}
&E_{i}(-t, -l ,{\bf x})=E_{i}^{L \dagger}(-t, -l ,{\bf x})+E_{i}^{R \dagger}(-t, -l ,{\bf x})+ E_{i}^{L}(-t, -l ,{\bf x})+E_{i}^R(-t, -l ,{\bf x}), \\
&B_{i}(-t, -l ,{\bf x})=B_{i}^{L \dagger}(-t, -l ,{\bf x})+ B_{i}^{R \dagger}(-t, -l ,{\bf x})+B_{i}^{L}(-t, -l ,{\bf x})+B_{i}^R(-t, -l ,{\bf x}), \\
\end{split}
\ee
where the subsystem $A$ is $x^1\ge 0$. The ground states for the left and light moving modes are defined by
\be 
\begin{split}
&E_{i}^{L, R}(-t, -l ,{\bf x}) \ket{0}_{L, R}=B_{i}^{L, R}(-t, -l ,{\bf x})\ket{0}_{L, R}=0, \\
&\ket{0}=\ket{0}_L \otimes \ket{0}_R. \\ 
\end{split}
\ee

The algebra which quasi-particles obey can be given by
\be
\begin{split}
&\left[E^L_i(-t, -l ,{\bf x}), E^{L \dagger}_j (-t, -l ,{\bf x})\right]=F^{(n)}_{EiEi}(\theta_1-\theta_2)\delta_{ij}, \\
&\left[E^R_i(-t, -l ,{\bf x}), E^{R \dagger}_j (-t, -l ,{\bf x})\right]=F^{(n)}_{EiEi}(\theta_1-\theta_2+2\pi)\delta_{ij}, \\
&\left[B^L_i(-t, -l ,{\bf x}), B^{L \dagger}_j (-t, -l ,{\bf x})\right]=F^{(n)}_{BiBi}(\theta_1-\theta_2)\delta_{ij}, \\
&\left[B^R_i(-t, -l ,{\bf x}), B^{R \dagger}_j (-t, -l ,{\bf x})\right]=F^{(n)}_{BiBi}(\theta_1-\theta_2+2\pi)\delta_{ij}, \\
&\left[E^L_2(-t, -l ,{\bf x}), B^{L \dagger}_3 (-t, -l ,{\bf x})\right]=F^{(n)}_{E2B3}(\theta_1-\theta_2), \\
&\left[E^R_2(-t, -l ,{\bf x}), B^{R \dagger}_3 (-t, -l ,{\bf x})\right]=F^{(n)}_{E2B3}(\theta_1-\theta_2+2\pi), \\
&\left[E^L_3(-t, -l ,{\bf x}), B^{L \dagger}_2 (-t, -l ,{\bf x})\right]=F^{(n)}_{E3B2}(\theta_1-\theta_2), \\
&\left[E^R_3(-t, -l ,{\bf x}), B^{R \dagger}_2 (-t, -l ,{\bf x})\right]=F^{(n)}_{E3B2}(\theta_1-\theta_2+2\pi), \\
\end{split}
\ee


\subsection{$6$ dimensional Maxwell Theory}

\subsubsection{The propagators}
The analytic continued Green's functions on $\Sigma_n$ are defined by
\begin{align}
	\langle F_{i j}(\theta) F_{l m}(\theta ') \rangle &= F_{F_{i j} F_{l m}}^{(n)} (\theta - \theta')
\end{align}
In the $\epsilon \rightarrow 0$ limit, their leading terms are as follows. 

For the case of $n=1$ in $t>0$, 
\begin{align}
	\begin{split}
		F_{F_{0i}F_{0i}}^{(1)} (\theta_1 - \theta_2) &= \frac{1}{16 \pi^3 \epsilon^6},\\
		F_{F_{ij}F_{ij}}^{(1)} (\theta_1 - \theta_2) &= \frac{1}{32 \pi^3 \epsilon^6} ~(i,j \neq 0).
	\end{split} \label{eq:d6propn1}
\end{align}

For the case of $n \ge 2$, if $l>t>0$ they are the same as in (\ref{eq:d6propn1}).
In $t \ge l$, they are as follows, with $i,j=2,3,4,5$ and $i \neq j$:
\begin{align}
	\begin{split}
		F_{F_{01} F_{01}}^{(n)}(\theta_1 - \theta_2) &= \frac{1}{256 \pi^3} \frac{(t+l)^3 (3l^2 - 9lt + 8t^2)}{t^5 \epsilon^6},\\
		F_{F_{01} F_{01}}^{(n)}(\theta_1 - \theta_2 + 2 \pi) &= \frac{1}{256 \pi^3} \frac{(t-l)^3 (3l^2 + 9lt + 8t^2)}{t^5 \epsilon^6},\\
		F_{F_{0i} F_{0i}}^{(n)}(\theta_1 - \theta_2) &= \frac{1}{1024 \pi^3} \frac{(l+t)^2 (32t^3 - 19lt^2 + 6l^2 t -3l^3)}{t^5 \epsilon^6}, \\
		F_{F_{0i} F_{0i}}^{(n)}(\theta_1 - \theta_2 + 2 \pi) &= \frac{1}{1024 \pi^3} \frac{32t^5 - 45lt^4 + 10l^3 t^2 + 3l^5}{t^5 \epsilon^6},\\
		F_{F_{1i} F_{1i}}^{(n)}(\theta_1 - \theta_2 ) &= \frac{1}{1024 \pi^3} \frac{16 t^5 + 15 lt^4 + 10 l^3 t^2 - 9l^5}{t^5 \epsilon^6}, \\
		F_{F_{1i} F_{1i}}^{(n)}(\theta_1 - \theta_2 + 2 \pi) &= \frac{1}{1024 \pi^3} \frac{16t^5 - 15t^4 l - 10t^2 l^3 + 9l^5}{t^5 \epsilon^6}, \\
		F_{F_{ij} F_{ij}}^{(n)}(\theta_1 - \theta_2) &= \frac{1}{512 \pi^3} \frac{(t+l)^3(3l^2 - 9lt + 8t^2)}{t^5 \epsilon^6}, \\
		F_{F_{ij} F_{ij}}^{(n)}(\theta_1 - \theta_2 + 2 \pi) &= \frac{1}{512 \pi^3} \frac{(t-l)^3 (8t^2 + 9lt + 3l^2)}{t^5 \epsilon^6}, \\
		F_{F_{0i} F_{1i}}^{(n)}(\theta_1 - \theta_2) = F_{F_{1i} F_{0i}}^{(n)}(\theta_1 - \theta_2) &= \frac{15}{1024 \pi^3} \frac{(t^2 - l^2)^2 }{t^4 \epsilon^6}, \\
		F_{F_{0i} F_{1i}}^{(n)}(\theta_1 - \theta_2 + 2 \pi) = F_{F_{1i} F_{0i}}^{(n)}(\theta_1 - \theta_2 + 2 \pi) &= - \frac{15}{1024 \pi^3} \frac{(t^2 -l^2)^2}{t^4 \epsilon^6}.
	\end{split} \label{6dMaxwellProp}
\end{align}

They have the property $F_{IJ}^{(n)} (\theta) = F_{IJ}^{(n)} (- \theta)$, and
due to the periodicity of $n$-sheeted Riemann surface, they all satisfy $F_{IJ}^{(n)} (\theta) = F_{IJ}^{(n)} (\theta \pm 2 \pi n)$.

They are related as, 
\begin{align}
	\begin{split}
		F_{F_{ij}F_{ij}}^{(n)}(\theta_1 - \theta_2) + F_{F_{ij} F_{ij}}^{(n)}(\theta_1 - \theta_2 + 2 \pi) &= F_{F_{ij} F_{ij}}^{(1)}(\theta_1 - \theta_2),\\
		F_{F_{0i} F_{1i}}^{(n)}(\theta_1 - \theta_2) + F_{F_{0i} F_{1i}}^{(n)}(\theta_1 - \theta_2 + 2 \pi) &= 0,\\
		F_{F_{1i} F_{0i}}^{(n)}(\theta_1 - \theta_2) + F_{F_{1i} F_{0i}}^{(n)}(\theta_1 - \theta_2 + 2 \pi) &= 0,
	\end{split}
\end{align}
where $n \geq 2$, $i,j=2,3,4,5$.

\subsubsection{The commutation relation}
The operators $F_{ij}(-t, -l, \textbf{x})$ are decomposed into left and right moving modes as 
\begin{align}
	F_{ij}(-t,-l,\textbf{x}) &= F_{ij}^{L\dagger}(-t,-l,\textbf{x}) + F_{ij}^{R\dagger}(-t,-l,\textbf{x}) + F_{ij}^{L}(-t,-l,\textbf{x}) + F_{ij}^{R}(-t,-l,\textbf{x})
\end{align}
where the subsystem $A$ is $x^1 \geq 0$. The ground state for left and right moving modes are defined as
\begin{align}
	\begin{split}
		&F_{ij}^{L}(-t,-l,\textbf{x}) |0 \rangle_L = F_{ij}^{R}(-t,-l,\textbf{x}) |0 \rangle_R = 0,\\
		&|0\rangle := |0 \rangle_L \otimes |0 \rangle_R.
	\end{split}
\end{align}

The algebra which the quasi-particles obey are 
\begin{align}
	\begin{split}
		[ F_{ij}^{L}(-t,-l,\textbf{x}), F_{lm}^{L\dagger}(-t,-l,\textbf{x}) ] &= F^{(n)}_{F_{ij} F_{lm}}(\theta_1 - \theta_2), \\
		[ F_{ij}^{R}(-t,-l,\textbf{x}), F_{lm}^{R\dagger}(-t,-l,\textbf{x}) ] &= F^{(n)}_{F_{ij} F_{lm}}(\theta_1 - \theta_2 + 2 \pi), 
	\end{split}
\end{align}
where the ones not on the list (\ref{6dMaxwellProp}) are zero.

\end{document}